\documentclass[%
aps,
prb,
preprint,
 showpacs,
 showkeys,
 amsmath,amssymb,amsfonts,
superscriptaddress,
floatfix,
longbibliography
]{revtex4-1}

\usepackage{graphicx}
\usepackage{tabularx}
\usepackage{dcolumn}
\usepackage{bm}
\usepackage{natbib}
\usepackage{ulem}

\usepackage{epsfig}
\usepackage{graphicx}
\usepackage{epstopdf}
\usepackage{subfigure}
\usepackage{xcolor}
\usepackage{tikz}

\usepackage{multirow} 
\usepackage{booktabs} 

\newcommand{\be}{\begin{equation}}
\newcommand{\ee}{\end{equation}}
\newcommand{\bI}{\boldsymbol{I}}
\newcommand{\bD}{\boldsymbol{D}}

\newcommand{\bu}{\boldsymbol{u}}
\newcommand{\bnabla}{\boldsymbol{\nabla}}
\newcommand{\bsig}{\boldsymbol{\sigma}}
\newcommand{\btau}{\boldsymbol{\tau}}

\newcommand{\btauD}{\tau}

\newcommand{\pard}[2]{\frac{\partial #1}{\partial #2}}

\def\half{{\mbox{$\frac12$}}}
\def\quar{{\mbox{$\frac14$}}}
\def\v0{{V}}
\def\f0{{F}}
\def\ttt0{{T_{\theta\theta}}}
\def\trt0{{T_{r\theta}}}
\def\ftt{{f_{\theta\theta}}}
\def\frt{{f_{r\theta}}}

\def\di{{\rm d}}

\def\Bi{{B}}
\def\Wi{{W}}

\begin{document}

\title{Inertialess instability of elasto-viscoplastic circular Couette flow}

\author{James D. Shemilt}
\affiliation{Department of Mathematics, University of British Columbia
  Vancouver, BC, V6T 1Z2, Canada}
   
\author{Neil J. Balmforth}
\affiliation{Department of Mathematics, University of British Columbia
  Vancouver, BC, V6T 1Z2, Canada}

\author{Duncan R. Hewitt}
\affiliation{Department of Applied Mathematics and Theoretical Physics,
University of Cambridge, Wilberforce Road, Cambridge CB3 0WA, UK}

\date{\today}

\begin{abstract}
  An exploration is presented of two-dimensional, circular Couette
  flow of an elasto-viscoplastic fluid described by Saramito's
  (Bingham) constitutive law.
  Axisymmetrical base states are constructed. As has been noted previously,
  such states are not unique, with stresses over any plugs
  depending on the initial polymer stress, and the shear rate
  and stress may jump discontinuously across the yield surfaces.
  We show that base states with continuous stress profiles are
  linearly unstable at zero Reynolds number when they contain regions with relatively
  low shear rate. Unstable modes become stronger and localized
  to the yield surfaces as the azimuthal wavenumber increases.
  We report numerical simulations exploring the nonlinear
  dynamics of the instability. The linear unstable modes
  saturate at finite amplitude to create spatio-temporally
  complicated states. 
  Polymer stress diffusion is added to ease these numerical simulations,
  eliminating normal-stress discontinuities and instability at the
  shortest wavelengths.
  Instability and transition to complex dynamics occur whenever
  the base flow contains regions with relatively low shear rate. 

\end{abstract}

\maketitle

\numberwithin{equation}{section}

\section{Introduction}

Circular Couette flows are often useful, experimentally
realizable configurations in which to explore
the consequences of fluid rheology. For example,
a yield-stress fluid remains fully yielded in a narrow
annular gap, whereas the fluid plugs up near the
outer wall in wider annuli\cite{bird83}.
By contrast, for simple viscoelastic liquids like
the Oldroyd-B fluid, the axisymmetrical velocity profile is identical
to that for a Newtonian fluid \cite{bird,larson90}.
We focus in the current paper on elasto-viscoplastic
fluids\cite{fraggedakis16,coussot25},
which have generated much recent interest as a way to describe 
the impact on flow dynamics
of fluid elasticity acting in concert with a yield
stress\cite{franca2024,franca26,erken_2022_elastoviscoplastic,
sanjay21,balasubramanian24,mosch,kordalis26,fraggedakis16,chaparian20,
zakeri25,vpthread}. 

The circular Couette flow of an elasto-viscoplastic fluid
has been studied previously by 
Cheddadi {\it et al.}\cite{cheddadi08,cheddadi12} as a test problem
for a popular constitutive law proposed by
Saramito\cite{saramito07,saramito09}.
For axisymmetrical flows,
Cheddadi {\it et al.} showed that the stress profiles over any plug
were not unique for Saramito's model,
and that stress jumps could be generated at the
yield surfaces. They did not, however, explore whether such flows
could be linearly unstable towards non-axisymmetrical perturbations.

The purpose of the present article is to
advance beyond axisymmetrical flow and
determine whether elasto-viscoplastic circular Couette flow
are linearly unstable. We focus on inertia-less,
two-dimensional flow. Instabilities requiring a finite Reynolds
number or the axial spatial structure for a Taylor-Couette instability
are then ruled out ({\it e.g.} Refs.~\onlinecite{larson90,landry}).
Instead, we search for instability prompted by the yield stress,
motivated by recent results on elasto-viscoplastic Poiseuille
flow\cite{evpoiy},
where a novel linear instability was discovered connected to
the presence of the plug surrounding the centreline of 
the unperturbed base flow.
Part of our motivation is to establish that this instability
is not an exclusive feature of pressure-driven flow, but can also
appear in the wall-driven circular Couette configuration.
Moreover, the annular geometry provides a convenient setting in which to probe the link between instability and the
presence of unyielded plugs. 
In circular Couette flow, a plug can be introduced or eliminated
into the base state by tuning
the gap between the cylinders or the yield stress; in Poiseuille flow, by contrast, base-state plugs around the centre of the channel are unavoidable. 

In the Poiseuille problem, numerical simulations show that the linear instability develops into a spatio-temporally complicated flow state, with fluctuations mostly concentrated around the yield surface of the central plug\cite{evturbo}. 
In a similar manner, a further goal of the current article is to use numerical simulations to explore the nonlinear dynamics of any linear instability
for circular Couette flow. One practical reason to consider this configuration is that, being a standard rheometric
flow, it lends itself much more naturally to complementary
experimental studies than the pressure-driven channel problem, although here our approach remains purely theoretical.

In \S \ref{sec:form}, we set out the mathematical formulation
of the 
Couette problem for a fluid described by Saramito's
visco-elastic Bingham model\cite{saramito07}
and cast the equations into a convenient
dimensionless form. Setting aside any dependence on angle,
we then explore axisymmetrical base states
and determine their final steady form
in \S \ref{sec:BS}.
The base states are varied in view of the different stress
profiles that can be frozen into any plug and the stress jumps
that can arise across their yield surfaces. However, if 
the inner cylinder is rotated initially at a relatively large rate,
the entire fluid annulus becomes fully yielded at early times.
Stress relaxation at later times then leads to steady states in
which the fluid remains either above or at the yield stress.
Following Walton \& Bittleston\cite{walton_axial_1991}, we refer to
the regions held at the yield stress as ``pseudo-plugs''.
Critically, steady base states with pseudo-plugs
have continuous stress profiles.
In \S \ref{sec:LS}, we consider
the linear stability of such steady base states, their continuity
permitting a standard type of normal-mode analysis.
As in elasto-viscoplastic Poiseuille flow, we discover that
inertia-less flow is linearly unstable,
and in \S \ref{sec:nonlin} we interrogate the nonlinear dynamics, both 
for base states developing stress discontinuities and for those with pseudo-plugs. 
Surprisingly, we also find that a zero-shear rate plug
or pseudo-plug is not an essential ingredient for instability. Rather, fully yielded base states with a low but non-zero shear
rate near the outer wall can also be unstable; again, we explore the resulting nonlinear dynamics.

\section{Dimensionless governing equations}\label{sec:form}

We consider two-dimensional, inertialess (zero Reynolds number) flow of an
elasto-viscoplastic fluid described by Saramito's
Bingham model,\cite{saramito07}
in the two-dimensional annular domain $[R_1,R_2]$.
We quote (below) the governing equations in a dimensionless form
in which lengths are scaled by the inner radius $R_1$, velocities by a
characteristic velocity of the inner cylinder wall $U$,
time by $R_1/U$, and
pressure and stress by $\mu U/R_1$, where $\mu=\mu_s+\mu_p$ is the
total zero-shear viscosity, combining solvent and polymer
components, $\mu_s$ and $\mu_p$, respectively.
Four key dimensionless groups arise: the Weissenberg number, Bingham
number, viscosity ratio and outer cylinder position, defined by
\begin{equation}
  \Wi=\frac{\lambda U}{R_1},\quad
  \Bi=\frac{\tau_y R_1}{\mu U},\quad
  \beta=\frac{\mu_s}{\mu_s+\mu_p},\quad
  R = \frac{R_2}{R_1}.
\end{equation}
Here, $\lambda$ is the polymer relaxation time and
$\tau_y$ is the yield stress.


In Saramito's constitutive law, the total Cauchy stress is broken down
into a pressure, a viscous Newtonian (`solvent') contribution and the polymer
 extra stress, $\btau$:
\begin{equation}
  \bsig = -p\,\bI + 2\beta\bD + \btau,
\end{equation}
where $\bI$ is the identity and the
deformation-rate tensor $\bD=\tfrac{1}{2}(\bnabla\bu+(\bnabla\bu)^T)$
has components,
\begin{equation}
  \begin{aligned}
  D_{rr}&=\pard{u}{r},\qquad
  D_{\theta\theta}=\frac{u}{r}+\frac{1}{r}\pard{v}{\theta},\\
  D_{r\theta}&=\frac{1}{2}\!\left(\pard{v}{r}-\frac{v}{r}
    +\frac{1}{r}\pard{u}{\theta}\right).
  \end{aligned}
\end{equation}
In circular polar coordinates $(r,\theta)$,
with velocity $(u,v)$,
the governing equations comprise continuity,
and the component forms of force balance and the
constitutive law:
\begin{equation}
  \frac{1}{r}\pard{(ru)}{r}+\frac{1}{r}\pard{v}{\theta}=0.
  \label{eq:gv1}
\end{equation}
\begin{equation}
  -\pard{p}{r}
  +\beta\!\left[\nabla^2 u-\frac{u}{r^2}-\frac{2}{r^2}\pard{v}{\theta}\right]
  +\pard{\tau_{rr}}{r}+\frac{\tau_{rr}-\tau_{\theta\theta}}{r}
  +\frac{1}{r}\pard{\tau_{r\theta}}{\theta}=0.
\end{equation}
\begin{equation}
  -\frac{1}{r}\pard{p}{\theta}
  +\beta\!\left[\nabla^2 v-\frac{v}{r^2}+\frac{2}{r^2}\pard{u}{\theta}\right]
  +\frac{1}{r^2}\pard{(r^2\tau_{r\theta})}{r}
  +\frac{1}{r}\pard{\tau_{\theta\theta}}{\theta}=0.
\end{equation}
\begin{align}
  &\Wi\!\left(\frac{D\tau_{rr}}{Dt}
    -2\pard{u}{r}\,\tau_{rr}
    -\frac{2}{r}\pard{u}{\theta}\tau_{r\theta}\right)
    +f\,\tau_{rr} = 2(1-\beta)
    \pard{u}{r},
\label{eq:comp_rr}\\
  &\Wi\!\left[\frac{D\tau_{r\theta}}{Dt}
    - r\tau_{rr}\pard{}{r}\left(\frac{v}{r}\right)
    -\frac{1}{r}\pard{u}{\theta}\tau_{\theta\theta} 
    \right]
  + f\,\tau_{r\theta} 
 = (1-\beta) 
  \left(\pard{v}{r}-\frac{v}{r}+\frac{1}{r}\pard{u}{\theta}\right),
  \label{eq:comp_rt}\\
  &\Wi\!\left[\frac{D\tau_{\theta\theta}}{Dt}
  -2r\tau_{r\theta}\pard{}{r}\left(\frac{v}{r}\right) 
  +2\pard{u}{r}\tau_{\theta\theta}
  \right]
  + f\,\tau_{\theta\theta} 
  = 2(1-\beta) 
  \!\left(\frac{u}{r}+\frac{1}{r}\pard{v}{\theta}\right),
  \label{eq:comp_tt}
\end{align}
where 
\be
  f \equiv \max\!\left(0,1-\frac{\Bi}{\btauD}\right),
  \label{eq:fdef}
\ee
the second invariant of the polymer stress is
\be
  \btauD 
  = \sqrt{\quar (\tau_{rr}-\tau_{\theta\theta})^2+\tau_{r\theta}^2}
  \label{eq:tauD_norm}
\ee
and
\begin{equation}
  \frac{D}{Dt}=\pard{}{t}+u\pard{}{r}+\frac{v}{r}\pard{}{\theta}
    \label{eq:gv9}
\end{equation}
denotes the {\sl scalar} material derivative.

As a word of caution, we note that in some earlier papers
({\it e.g.} Ref.~\onlinecite{cheddadi08}), 
the scalar material derivative appears to have been
introduced into the constitutive law, even though
the material derivative of a tensor is demanded. This
leads to a number of erroneous terms with factors of
$\Wi$ on the left-hand sides of \eqref{eq:comp_rr}-\eqref{eq:comp_tt},
which are corrected above ({\it cf.} Appendix A of Cheddadi
{\it et al.}\cite{cheddadi08}, for the axisymmetrical version
of the equations, but which was later corrected by the same
authors\cite{cheddadi09,cheddadi12}). One consequence of the erroneous terms
is that, in the Oldroyd-B limit of the model, 
steady axisymmetrical flow does not reduce to a Newtonian shear flow
(whereas in fact it should\cite{larson90}).

No-slip boundary conditions on the cylinders impose
\begin{align}
  u&=0,\quad v=\Omega_0(t) \;\text{ at }r=1;\label{eq:ubc1}\\
  u&=v=0 \;\text{ at }r=R,\label{eq:ubcR}
\end{align}
where $\Omega_0(t)$ represents the prescribed (dimensionless)
rotation rate of the inner cylinder. We will consider both situations in which $\Omega_0$ is a constant, and those in which it evolves over time, imparting a different stress history to the fluid in the annulus. In view of the scaling of the
problem, the final or average value of $\Omega_0(t)$ can be
set to unity.

\section{Axisymmetric base states}\label{sec:BS}

For purely azimuthal flow with $u=0$,
we take $v=v(r,t)=r\Omega(r,t)$ and
$\boldsymbol{\tau}=\boldsymbol{T}(r,t)$.
The material derivative then reduces to
$D/Dt=\partial/\partial t$, and the only
non-zero rate-of-deformation component is the shear rate,
\begin{equation}
  \dot\gamma \equiv \pard{v}{r}-\frac{v}{r}
 \equiv r \pard{\Omega}{r}.
\end{equation}
Continuity is satisfied identically.
The force balance and constitutive equations reduce to
\begin{align}
  -\pard{p}{r}
  +\frac{1}{r}\pard{(r T_{rr})}{r}
  -\frac{T_{\theta\theta}}{r} &= 0,
  \label{eq:mom_r_az}\\
  \frac{1}{r^2}\frac{\partial}{\partial r}\!
  \left[r^2\!\left(\beta\dot\gamma+T_{r\theta}
    \right)\right]&=0,
  \label{eq:mom_t_az} \\
  \Wi\,\pard{T_{rr}}{t}
  + F\,T_{rr} &= 0,
  \label{eq:az_rr}\\
  \Wi\!\left(\pard{T_{r\theta}}{t}
    - \dot\gamma\,T_{rr}\right)
  + F\, T_{r\theta}
  &= (1-\beta)\dot\gamma,
  \label{eq:az_rt}\\
  \Wi\!\left(\pard{T_{\theta\theta}}{t}
    - 2\dot\gamma\,T_{r\theta}\right)
  + F\,T_{\theta\theta} &= 0,
  \label{eq:az_tt}\\
  F = \max\left(0,1-\frac{\Bi}{T}\right),\quad
  T &= \sqrt{T_{r\theta}^2+\quar (T_{rr}-T_{\theta\theta})^2}.
\end{align}
Equation \eqref{eq:az_rr} implies that $T_{rr}$ is unforced. Hence
this stress component
decays to zero over the yielded regions ($F>0$) on the Weissenberg timescale,
while being frozen at its current value
in the unyielded region ($F=0$) 
To simplify matters, we assume
that $T_{rr}(r,0)=0$, and therefore omit the radial normal stress.

From \eqref{eq:mom_t_az}, we arrive at the local torque balance,
\begin{equation}
  \beta\dot\gamma + T_{r\theta} = \frac{C}{r^2},
  \label{eq:torque}
\end{equation}
where the constant $C$ is the torque per unit length on either
cylinder. Given the boundary conditions on the cylinders,
a radial integral of \eqref{eq:torque}
furnishes the constraint
\be
\int_1^R \frac{T_{r\theta}}{r}\di r - \beta \Omega_0 
= \half C \left(1 - R^{-2}\right)
.
\label{coni}
\ee
Equations \eqref{eq:az_rt} and \eqref{eq:az_tt} now constitute a
pair of ordinary differential equations to solve at each
radius $r$, given the shear rate from \eqref{eq:torque}
and the additional constraint \eqref{coni} which dictates
$C$ once the velocity conditions $\Omega(1,t)=\Omega_0(t)$
and $\Omega(R,t)=0$ are imposed.
Note that \eqref{eq:mom_r_az} serves only to
dictate the radial pressure gradient.

\subsection{Steady states}

After setting $\partial/\partial t=0$, $T_{rr}=0$ and
$\Omega_0=1$,
the constitutive relations~\eqref{eq:az_rt}--\eqref{eq:az_tt}
reduce to
\begin{align}
  F \, T_{r\theta} &= (1-\beta)\,\dot\gamma, \label{ss1} \\
  F \, T_{\theta\theta} &= 2\Wi\,\dot\gamma\,T_{r\theta} , \label{ss2} \\
  F &= \max\left(0,1-\frac{\Bi}{T}\right),\quad
  T = \sqrt{T_{r\theta}^2+\quar T_{\theta\theta}^2}.
    \label{ss3}
\end{align}

\subsubsection{Oldroyd-B limit ($\Bi=0$)}

When $\Bi=0$, $F\equiv 1$ and the
model reduces to the Oldroyd-B fluid. In this limit, the steady-state
solution reduces to
\begin{align}
  T_{r\theta} &= (1-\beta)\,\dot\gamma,
  \label{eq:OB_trt}\\
  T_{\theta\theta} &= 2(1-\beta)\Wi\,\dot\gamma^2,
  \label{eq:OB_ttt}\\
  \dot\gamma &= \frac{C}{r^2}.
  \label{eq:OB_torque}
\end{align}
Thus, the velocity profile is identical to the
Newtonian Couette solution,
\begin{equation}
  v = \frac{R^2-r^2}{r(R^2-1)}\qquad
  \left(C = -\frac{2 R^2}{R^2-1}\right)
  ,
  \label{eq:Newt}
\end{equation}
although the normal stress is non-trivial,
\begin{equation}
  T_{\theta\theta} = \frac{2(1-\beta)\Wi\, C^2}{r^4},
\end{equation}
and drives a radial pressure gradient through~\eqref{eq:mom_r_az}.

\subsubsection{elasto-viscoplastic cases}

Over the yielded regions ($F>0$), \eqref{ss1}-\eqref{ss3} simplify to
\begin{align}
    \dot\gamma &= \frac{F\,T_{r\theta}}{1-\beta},
  \label{eq:trt_steady} \\
  T_{\theta\theta} &= \frac{2\Wi\,T_{r\theta}^2}{1-\beta},
  \label{eq:ttt_steady} \\
    F &= 1-\frac{\Bi}{|T_{r\theta}|}
    \left[1+\frac{\Wi^2T_{r\theta}^2}{(1-\beta)^2}\right]^{-1/2}.
    \label{eq:ycs} 
\end{align}
After inserting \eqref{eq:trt_steady} and \eqref{eq:ycs} and rearranging,
the torque balance \eqref{eq:torque} now boils down to an algebraic
equation for the shear stress:
\begin{equation}
  \left[T_{r\theta}-\frac{(1-\beta)C_\infty}{r^2}\right]
  \!\sqrt{1+\frac{\Wi^2T_{r\theta}^2}{(1-\beta)^2}}
  = -\beta\Bi,
  \label{eq:trt_implicit}
\end{equation}
where $C_\infty$ is the steady state value of $C$. 
 
Over any unyielded region ($F=0$) the material undergoes rigid-body rotation
with $v=r\Omega_p$ and $\dot\gamma=0$, for some constant
rotation rate $\Omega_p$.
Here, torque balance~\eqref{eq:torque} indicates that
\be
T_{r\theta}=\frac{C_\infty}{r^2}
.
\label{Trts}
\ee
But with $F=\dot\gamma=0$, \eqref{ss1}--\eqref{ss2}
are automatically satisfied,
implying that the stress component $T_{\theta\theta}$ is not determined,
provided the yield condition $|\btauD|\le\Bi$ is met.
In other words, the normal stress over the plug
is dictated by the history of elastic and plastic
deformation\cite{cheddadi12}.
Given that $T>|T_{r\theta}|=|C_\infty|/r^2$, one expects
that there is always a yielded region against the inner cylinder, but
a plug may appear at the outer cylinder
(with $\Omega_p=0$) for sufficiently wide gaps.

To find the plug stress, one must therefore return to the
initial-value problem in \eqref{eq:az_rt} and \eqref{eq:az_tt}.
If the fluid {\sl never} yields at a particular radius, then
\eqref{eq:az_rt} and \eqref{eq:torque} imply that
\be
T_{r\theta}(r,t) = T_{r\theta}(r,0) e^{-\Gamma t} +
\frac{\Gamma}{r^2} \int_0^t e^{\Gamma (s-t)}C(s)\,\di s
,
\ee
where
\be
\Gamma  = \frac{1-\beta}{\beta\Wi},
\ee
which recovers the steady state in \eqref{Trts} for $t\to\infty$.
Moreover, if the initial stresses vanish, one can further show that
\be
T_{\theta\theta} \to \frac{WT_{r\theta}^2}{1-\beta}=
\frac{\Wi \; C_\infty^2}{r^4(1-\beta)}
\label{Ttts}
.
\ee
At the yield surface, $r=r_y$, the condition $T=\Bi$ now demands
that
\be
r_y^4 = \frac{C_\infty^2}{2\mathrm{\Bi}^2}
\left(1 + \sqrt{1 + \frac{\Wi^2\mathrm{\Bi}^2}{(1-\beta)^2}}\right).
\label{eq:ry}
\ee
The two relations in \eqref{Trts} and \eqref{Ttts} imply
certain limits for the polymer stress to the plugged
side of the yield surface. In general, however,
these limiting stresses do not agree with those from
the yielded side, given by
\eqref{eq:trt_implicit} and \eqref{eq:ttt_steady} with $r=r_y$.
A stress discontinuity must therefore appear,
the relaxed stress over yielded region being
inconsistent with the stress history over the plug.
The discontinuity implies further that $\dot\gamma$ is also not continuous
and there is a jump in shear rate.
These conclusions were reached previously by
Cheddadi {\it et al.}\cite{cheddadi08,cheddadi12}.

A notable special case arises when, in the final steady state,
the yield criterion $T=\Bi$ is met throughout the
unyielded region. As
the stress lies exactly on the yield surface rather than strictly
below it, we refer to this state of perfectly plastic
deformation as a pseudo-plug,
following the terminology of Walton \& Bittleston\cite{walton_axial_1991}.
The shear stress is once more given by \eqref{Trts}. However,
the marginal yield condition $T=\Bi$, now
fixes the normal stress as
\begin{equation}
  T_{\theta\theta}(r) = 2\sqrt{\Bi^2-\frac{C_\infty^2}{r^4}},
  \qquad r\in[r_p,R],
  \label{eq:pseudo_ttt}
\end{equation}
which now becomes continuous across the yield surface.
Moreover, from \eqref{eq:trt_implicit} and \eqref{Trts}, it follows
the border of the pseudo-plug $r=r_p$ is given by
\begin{equation}
%
r_p^4 = \frac{C_\infty^2}{2\mathrm{\Bi}^2}
        \left(1 + \sqrt{1 + \frac{4\Wi^2\Bi^2}{(1-\beta)^2}}\right).
    \label{eq:ry_explicit}
\end{equation}

\subsection{Sample base states}

\begin{figure*}
  \centering
	\includegraphics[width=\linewidth]{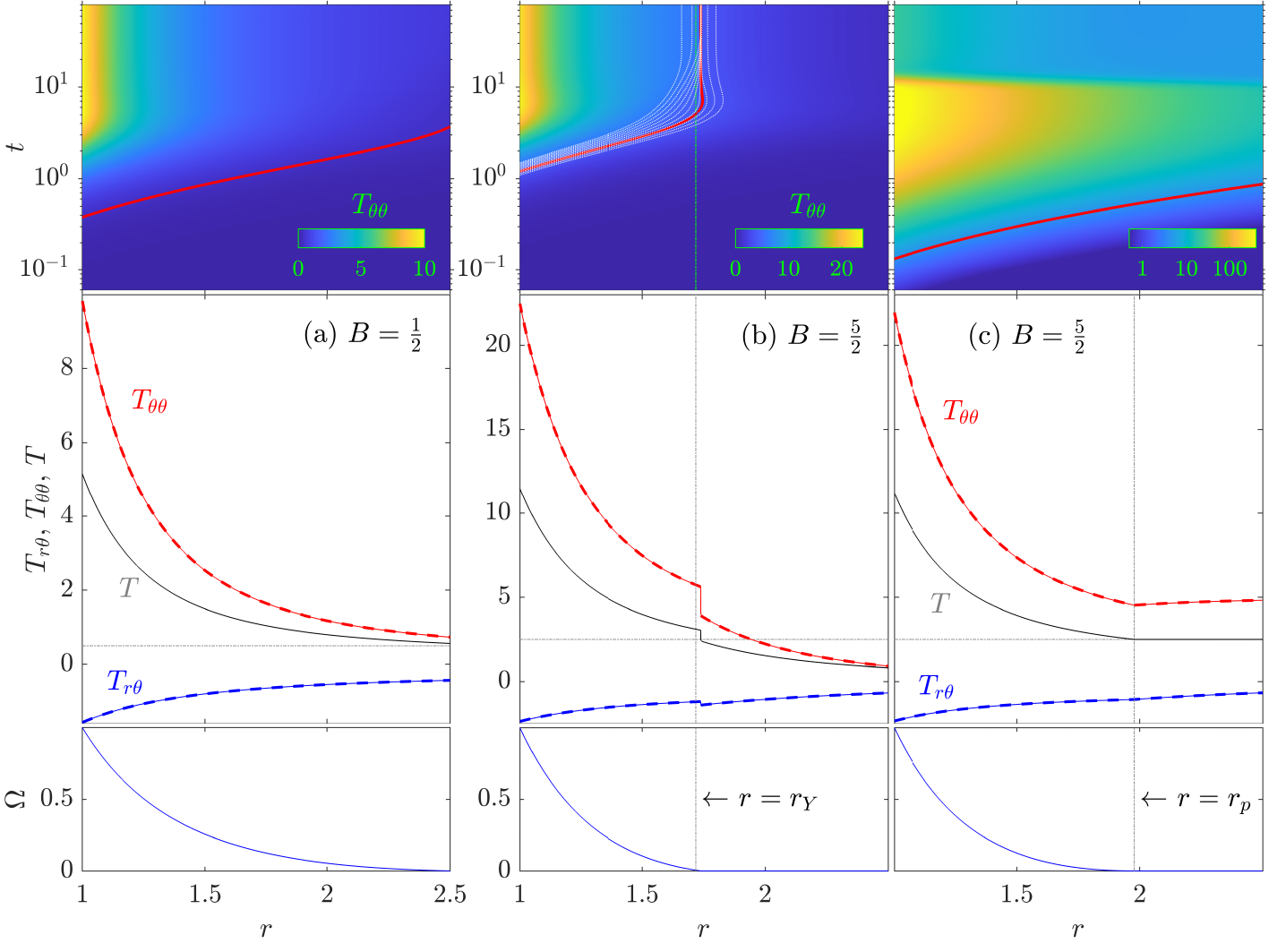}
	\caption{Solutions of the initial-value problem for
          base states with $(\Wi,\beta,R)=(1,\frac12,\frac52)$
          and the values of $\Bi$ indicated. In (a) and (b), the inner cylinder
          has rotation rate $\Omega_0=1$; for (c),
          $\Omega_0(t)$ is given by \eqref{Om0} with $A_0=4$.
          The top row of panels shows $\tau_{\theta\theta}(r,t)$ as
          density plots over the $(r,t)-$plane; the red contour shows the
          yield surface. The middle row of panels display the
          final snapshots of the polymer stress components and invariant;
          the dashed lines indicate the expected profiles from
          \eqref{eq:trt_implicit} and \eqref{eq:ttt_steady} (in
          (a) and (b)), or 
          \eqref{Trts} and \eqref{Ttts} (in (b)), then
          \eqref{eq:pseudo_ttt} in (c).
          The yield positions from \eqref{eq:ry} or \eqref{eq:ry_explicit}
          are shown by the lighter dash-dotted lines in (b) and (c)
          (respectively), along with $\tau=\Bi$.
          For all these steady-state profiles, the
          final snapshot of $\tau_{r\theta}$ is used to compute
          $C_\infty$ from \eqref{coni}. In (b),          
          additional contours of constant
          $\tau_{\theta\theta}$ are shown by the dotted white lines
          in the density map. 
        }
	\label{fig:sarexplot}
\end{figure*}

Sample solutions to the initial-value problem for three different base states
are shown in figure \ref{fig:sarexplot}. In all three examples,
the initial polymer stress is taken to be zero, and we fix $R=\tfrac{5}{2}$ and
$\beta=\tfrac{1}{2}$.
In the first two cases, the inner cylinder rotates with rate $\Omega_0=1$.
For the first solution
in panel (a), the Bingham number is too small for a plug to form at the outer cylinder;
the polymer stress builds up due to the shear imposed by the inner
cylinder, yielding the fluid for times of order unity. Thereafter,
the stress distribution converges to the steady state given everywhere
by \eqref{eq:trt_implicit} and \eqref{coni} (shown by the dashed
lines in figure \ref{fig:sarexplot}(a)).

For the second example in panel (b), $\Bi$ is higher.
The polymer stress then does not grow sufficiently to yield the
fluid near the outer cylinder, and an outer stationary plug persists
there indefinitely. To either side of the yield surface,
$r=r_Y\approx1.718$, the stresses then relax to different states
and a stress jump arises. For $r<r_Y$, the stress profile is
again given by \eqref{eq:trt_implicit} and \eqref{coni};
for $r>r_Y$, the profile instead follows from
\eqref{Trts} and \eqref{Ttts}. These two pieces of the stress profile
become combined through \eqref{coni} to largely dictate $C$.
However, as can be seen in the upper panel of
figure \ref{fig:sarexplot}(b), during the evolution
a thin layer of fluid in $r_Y<r<1.738$ first yields before plugging back up
at later times. That layer therefore acquires a small amount of
unrecoverable plastic strain, rendering \eqref{Ttts} invalid
and introducing a minor discrepancy between the final
numerical solution and the prediction using that analytical profile
and \eqref{eq:ry}. We note that a similar base state with an outer plug could also be achieved by increasing $R$ at fixed $\Bi$, rather than increasing $\Bi$ at fixed $R$ as we did in (b).

For the final solution in figure \ref{fig:sarexplot}(c), the rotation rate of the
inner cylinder is not fixed but
given by
\be
\Omega_0(t)=1 + A_0 [1-\tanh(t-10)]
,
\label{Om0}
\ee
with $A_0=4$.
The cylinder therefore starts rotating with higher speed, before slowing
to unit speed for times around $t=10$. This ``pre-shear'' leads to
a polymer stress that builds initially, but then decays for
later times. The consequence of the initial elevation in stress
is that stresses uniformly decline at late times
from a fully yielded state. Crucially, a pseudo-plug
then forms against the outer cylinder where the stress remains
above the yield stress by an
arbitrarily small amount for all finite times.
The final, continuous stress profile then follows by combining
\eqref{eq:trt_implicit} and \eqref{eq:ttt_steady} with
\eqref{Trts} and \eqref{eq:pseudo_ttt},
as shown by the dashed lines 
in \ref{fig:sarexplot}(c).

\begin{figure}
  \centering
	\includegraphics[width=.95\linewidth]{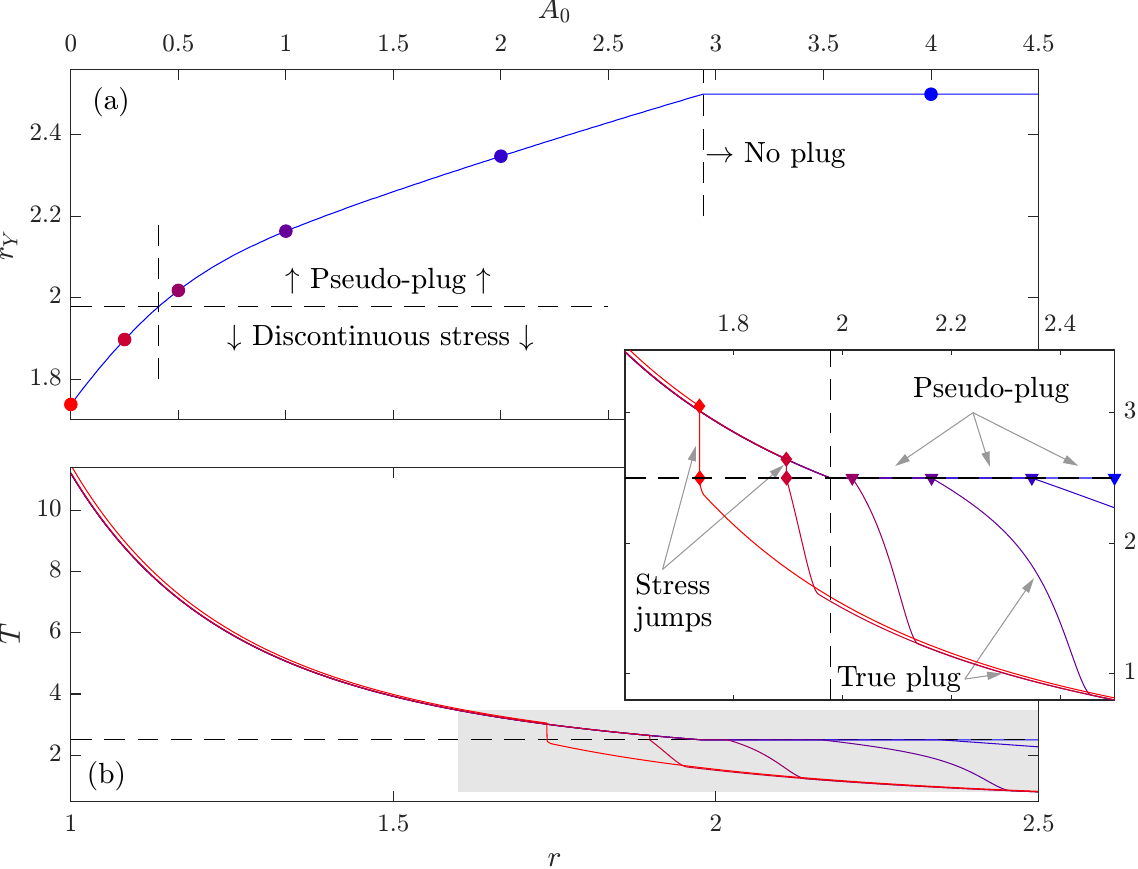}
	\caption{ (a) Yield radii, as a function of the pre-shear coefficient $A_0$, and (b) final stress-invariant profiles,
          for base states with $(\Bi,\Wi,\beta,R)=(\frac52,1,\frac12,\frac52)$
          and inner rotation rate \eqref{Om0}.
          The divisions between final stress states
          with discontinuities, pseudo-plugs or no plugs are
          indicated by the dashed lines in (a). In (b), the
          final profiles are shown for the values of $A_0$ indicated
          in (a); the dashed lines show $\tau=\Bi$ and the yield
          radius beyond which the stress profile is no longer
          discontinuous.
          The overlaid plot shows a magnification of the shaded region
          in (b); the
          stress discontinuities are indicated by diamonds
          and continuous yield points by triangles.
        }
	\label{fig:sarplugo}
\end{figure}

Further details of the effect of pre-shearing are shown in
figure \ref{fig:sarplugo}. Panel (a) illustrates how the final plug
radius depends on $A_0$. For $A_0\lesssim0.41$, a
plug is established with discontinuous stresses at its yield surface.
When $A_0\gtrsim0.41$, however, a pseudo-plug appears, sandwiched
between the yielded inner region and an outer true plug. The intervention
of the pseudo-plug renders the stress profile continuous
(see panel (b)). As $A_0$ is increased further, the outer true plug
becomes ever thinner, until it disappears entirely at
$A_0\approx2.94$. For larger initial pre-shears, the final
state then looks like that in figure \ref{fig:sarexplot}(c).

\begin{figure}
  \centering
	\includegraphics[width=.8\linewidth]{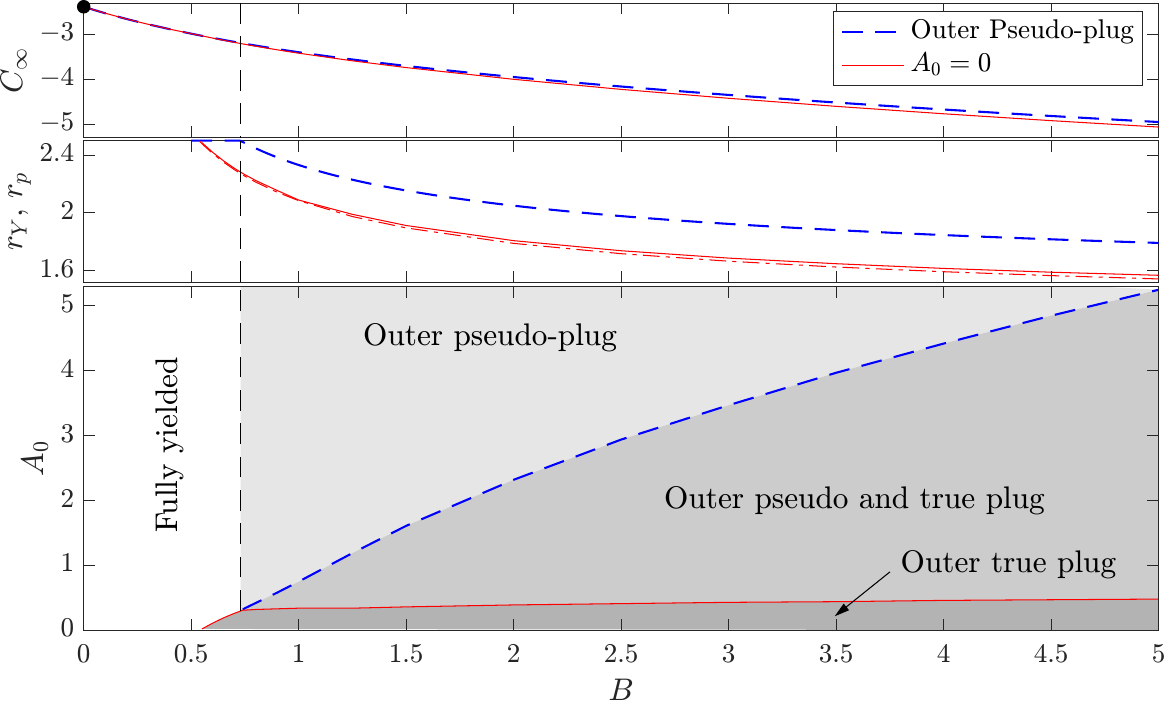}
	\caption{ (a) Cylinder torque $C_\infty$, (b) yield and pseudo-plug
          radii, and (c) regime diagram 
          for base states with $(\Wi,\beta,R)=(1,\frac12,\frac52)$,
          varying $\Bi$
          and inner rotation rate \eqref{Om0}.
          In (a) and (b),
          the (red) solid lines indicate the values for $A_0=0$
          and a base state with an outer true plug; the (blue) dashed
          lines show the results for an outer pseudo-plug. In (c),
          these lines show instead the borders of the regimes
          in which there is an outer true plug
          or pseudo-plug. The vertical dashed line indicates the
          yield stress below which no pseudo-plug arises for any $A_0$
          and the annulus becomes fully yielded for a sufficient pre-shear.
          The dot-dashed line in (b) indicates the approximation in
          \eqref{eq:ry}, given the corresponding torque from (a).
          The filled circle in (a) shows the Newtonian torque
          $C_\infty = -2R^2/(R^2-1)$.
        }
	\label{fig:regiplot}
\end{figure}

How these results depend on the yield stress $\Bi$ is indicated
in figure \ref{fig:regiplot}. For $\Bi\gtrsim0.73$, the base states
largely follow the pathway indicated in figure \ref{fig:sarplugo}:
discontinuous states with true plugs become continuous with the
appearance of a pseudo-plug on raising $A_0$; the true plug
then disappears for higher pre-shear, leaving an outer pseudo-plug.
For $\Bi\lesssim0.73$, however, the pseudo-plug does not appear, with
increasing pre-shear slowly narrowing the outer true plug until that
feature disappears, leaving a fully yielded annulus. The fully
yielded base state is the unique solution, regardless of pre-shear,
for $\Bi\lesssim0.55$. Note that, with $A_0=0$
and as in  figure \ref{fig:sarexplot}(b), the stress always overshoots
the yield stress beyond the final yield surface,
creating a layer with
unrecoverable plastic strain there, which relegates \eqref{eq:ry}
to an approximation for all $\Bi$.

\section{Linear stability}\label{sec:LS}

We explore the linear stability of the steady azimuthal base flows
of Section~\ref{sec:BS} towards non-axisymmetrical perturbations
as follows. To begin, we first restrict our attention to base states
without any plug, or with a pseudo-plug. For such states, the
stress profile remains continuous and we may ignore complications
associated with how one perturbs a stress discontinuity\cite{evpoiy}.
We next introduce normal-mode
perturbations of azimuthal wavenumber $m$ and complex growth rate
$\sigma$, {\it via}
\begin{equation}
  \begin{pmatrix} u\\ v\\ \tau_{rr}\\ \tau_{r\theta}\\ \tau_{\theta\theta}
  \end{pmatrix}
  =
  \begin{pmatrix} 0\\ r\Omega(r)\\ 0\\ \trt0(r)\\ \ttt0(r)
  \end{pmatrix}
  +
  \begin{pmatrix} \hat u(r)\\ \hat v(r)\\ \hat\tau_{rr}(r)\\
    \hat\tau_{r\theta}(r)\\ \hat\tau_{\theta\theta}(r)\end{pmatrix}
  e^{im\theta+\sigma t},
  \label{eq:LSA_ansatz}
\end{equation}
where $(\Omega(r),T_{r\theta}(r), T_{\theta,\theta}(r))$
denotes the steady base-state solution. The base state has shear rate
$r\Omega'$.
To account for the incompressibility condition, we define a streamfunction,
\be
\hat u=\frac{im}{r}\psi,
\qquad \hat v=-\psi',
\ee
then take the curl of the force balance equations
to eliminate the pressure. After
linearizing the constitutive equations
and defining
\be
D = \sigma + im\Omega + \frac{\f0}{\Wi},
\ee
we arrive at
\begin{align}
    \beta \left( \frac{1}{r}\frac{d}{dr}r\frac{d}{dr} -
  \frac{m^2}{r^2}\right)^2 \psi =&
  \frac{im}{r^2}\frac{d}{dr} \left[r (\hat\tau_{\theta\theta}  -
    \hat\tau_{rr})\right]
  + \left(\frac{d^2}{dr^2}+\frac{3}{r}\frac{d}{dr}+
  \frac{m^2}{r^2}\right)\hat\tau_{r\theta} ,
  \label{eq:app_vorticity}
  \\
  D\hat\tau_{rr}
    -\frac{2im}{r}\trt0\hat u
  =& \frac{2(1-\beta)}{\Wi}\hat u',
  \label{eq:LSA_rr}
  \\
  D\hat\tau_{r\theta}
    +\trt0'\hat u
    -r\Omega'\hat\tau_{rr}
  &-\frac{im}{r}\ttt0\hat u
  +\frac{\trt0\hat f}{\Wi}
  = \frac{1-\beta}{r\Wi}\left(r\hat v'-\hat v+im\hat u\right),
  \label{eq:LSA_rt}
\\
  D\hat\tau_{\theta\theta}
    +\ttt0'\hat u
    -2r\Omega'\hat\tau_{r\theta}
  &-2r\trt0\left(\frac{\hat v}{r}\right)'
    +2\ttt0\hat u'
  + \frac{\ttt0\hat f}{\Wi}
  = \frac{2(1-\beta)}{r\Wi}\left(\hat u+im\hat v\right),
\label{eq:LSA_tt}
\end{align}
where
\begin{equation}
  \hat f =
\left\{ \begin{array}{ll}
  \Bi
  \left[4\trt0\hat\tau_{r\theta}- \ttt0
    (\hat\tau_{rr}-\hat\tau_{\theta\theta})\right]/(4T^{\,3}),
  & T\geq\Bi , \\
  0, & T<\Bi.
  \end{array}\right.
  \label{eq:LSA_fhat}
\end{equation}
The no-slip conditions on the cylinders translates
to the boundary conditions,
\be
\psi(1)=\psi'(1)=\psi(R)=\psi'(R)=0.
\ee

\begin{figure*}
  \centering
	\includegraphics[width=\linewidth]{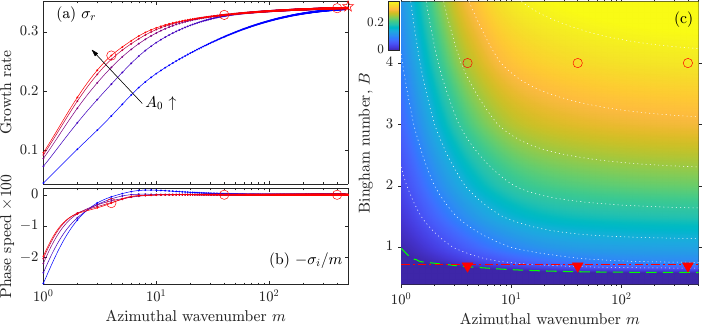}
	\caption{Growth rates $\sigma_r$ and phase speeds $-\sigma_i/m$
          of the most unstable normal mode
          for base states with $(\Wi,\beta,R)=(1,\frac12,\frac52)$
          and varying $\Bi$. In (a) and (b), growth rates
          and phase speeds are plotted against $m$ for $\Bi=4$
          for base states with $A_0=\{\frac12,1,2,3,5\}$
          (colour-coded from blue to red;
          {\it i.e.} for the base states with true outer plugs
          with reducing size, see figure \ref{fig:regiplot}).
          The star in (a) shows the predictions of the
          short-wavelength analysis of Appendix \ref{sec:appB}.
          In (c), growth rates for a wider suite of computations are
          shown as a density over the $(m,\Bi)-$plane. The dashed line
          shows the stability boundary, the dot-dashed line shows
          the value of $\Bi$ below which the plug disappears
          and the dotted lines show the contours $\sigma=0.05$, $0.1$, $0.15$,
          $...$, $0.35$. The circles and triangles show the cases for which
          the eigenfunctions are plotted in figures \ref{fig:greig} and
          \ref{fig:greigxx}, respectively.
        }
	\label{fig:grplot2}
\end{figure*}

\begin{figure*}
  \centering
	\includegraphics[width=\linewidth]{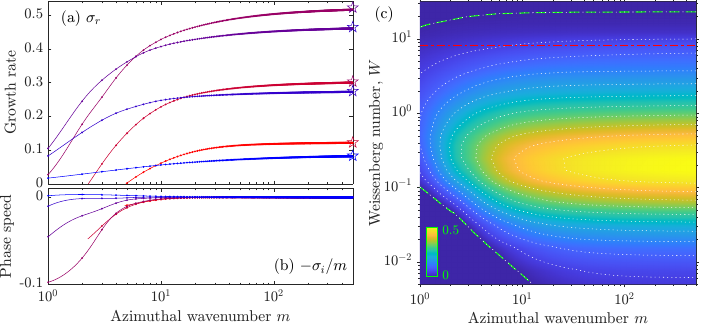}
	\caption{Growth rates $\sigma_r$ and phase speeds $-\sigma_i/m$
          of the most unstable normal mode
          for base states with $(\Bi,\beta,R)=(4,\frac12,\frac52)$
          and varying $\Wi$. In (a) and (b), growth rates
          and phase speeds are plotted against $m$ for
       $\Wi=\{\frac{1}{64},\frac{1}{32},\frac18,\frac12,\frac32,8\}$
          (color-coded, from red to blue).
          The stars in (a) show the predictions of the
          short-wavelength analysis of \S\ref{sec:appB}.
          In (c), growth rates for a wider suite of computations are
          shown as a density over the $(m,\Wi)-$plane. The dashed line
          shows the stability boundary, the dot-dashed line shows
          the value of $\Wi$ above which the plug disappears
          and the dotted lines show the contours $\sigma=0.05$, $0.1$, $0.15$,
          $...$, $0.5$.}
	\label{fig:grplot}
\end{figure*}

We solve the system in \eqref{eq:app_vorticity}-\eqref{eq:LSA_tt}
in two ways. First, we introduce a spatial grid
and use Chebyshev differentiation matrices to approximate radial
derivatives, turning the equations into a matrix eigenvalue problem.
Second, given a trial solution set by an eigenvalue/vector pair,
we solve the system as a boundary-value problem using MATLAB's
bvp4c. Sample growth rates $\sigma_r$  and phase speeds $-\sigma_i/m$
of the most unstable normal modes found in this fashion
are shown in figures \ref{fig:grplot2} and \ref{fig:grplot}.

In the first of these figures, we show results in panel (a) and (b)
for base states with $(\Bi,\Wi,\beta,R)=(4,1,\frac12,\frac52)$
and varying azimuthal wavenumber $m$. The choices of $R=2.5$ and $\beta=\tfrac{1}{2}$ do not have any particular significance, but are moderate values that we also use in numerical simulations in $\S \ref{sec:nonlin}$. Our focus is on exploring the effects of yield stress and elasticity by varying $\Bi$ and $\Wi$, but we also touch on the impacts of $\beta$ and $R$ in $\S\ref{sec:LSshort}$. As seen in figure \ref{fig:regiplot},
with the given parameter settings, a range of different base states are possible
with continuous stress profiles depending on the initial pre-shear $A_0$.
In figure \ref{fig:grplot2}(a,b), we show results for
$A_0=\{\frac12,1,2,3,5\}$, which correspond to states with
increasingly narrow true plugs against the outer cylinder; for
$A_0=5$, there is only an outer pseudo-plug.
For all these base state, normal modes are unstable, with the
state with the full pseudo-plug possessing the strongest instabilities.
As found in EVP Poiseuille flow\cite{evpoiy},
the growth rate increases with $m$, reaching a plateau for $m\gg1$.
Simultaneously, the phase speed decreases towards zero,
implying that mode patterns rotate increasingly slowly.

Note that the main difference between the base states with different
$A_0$ in figure \ref{fig:grplot2}(a,b) arises from
the switch in \eqref{eq:LSA_fhat}. Over the pseudo-plugs, the switch
allows the yield function
to become perturbed, the fluid being taken to be
slightly above the yield stress. Over the true plugs, however,
the switch sets $\hat f=0$. The variation in the growth rates
plotted in figure \ref{fig:grplot2}(a) therefore conveys a sense
of the possible range due to variations in the plug structure of the
base state.

A wider view of the stability characteristics is shown in
figure \ref{fig:grplot2}(c) and figure \ref{fig:grplot},
which present results for varying $m$ and suites of
bases states with varying $\Bi$ and $\Wi$, respectively,
for base states that are either fully yielded or have
outer pseudo-plugs (the pre-shear $A_0$ taken
sufficiently large).
For fixed $\Wi=1$ and varying yield stress,
instability is found for sufficiently large $\Bi$;
with fixed $\Bi=4$ and varying $\Wi$, instability is confined
to intermediate Weissenberg numbers, with base states becoming
stable for either sufficiently large or small $\Wi$.
Instability is invariably maximized at the highest azimuthal
wavenumbers, where phase speeds become small, with
long waves occasionally becoming stable.

In figure \ref{fig:grplot2}(c) and figure \ref{fig:grplot},
the unstable base states mostly have pseudo-plugs,
in line with the expectation that the instability is associated
with those regions\cite{evpoiy}. However, for smaller yield
stress in figure \ref{fig:grplot2}(c),
or for higher Weissenberg number in figure \ref{fig:grplot},
the pseudo-plug disappears (for $\Bi\lesssim0.73$ or
$\Wi\gtrsim8.22$, respectively). Yet,
as illustrated by the stability boundary also indicated
in the these figures, base states remain unstable 
until $\Bi$ falls below $\Bi\approx0.6$ or $\Wi$
becomes larger than about $23.2$.
In other words, even though there is no longer any
pseudo-plug (over the region between the red dot-dashed and green dashed
lines), the relatively low shear rates near the outer cylinder
are sufficient to drive instability. This result is somewhat reminiscent of the instability found previously for shear-thinning viscoelastic fluids in 
channel flow\cite{wilson99,wilson15,castillo18,bodiguel,poole16}, where there also is no plug.

\begin{figure}
  \centering
	\includegraphics[width=\linewidth]{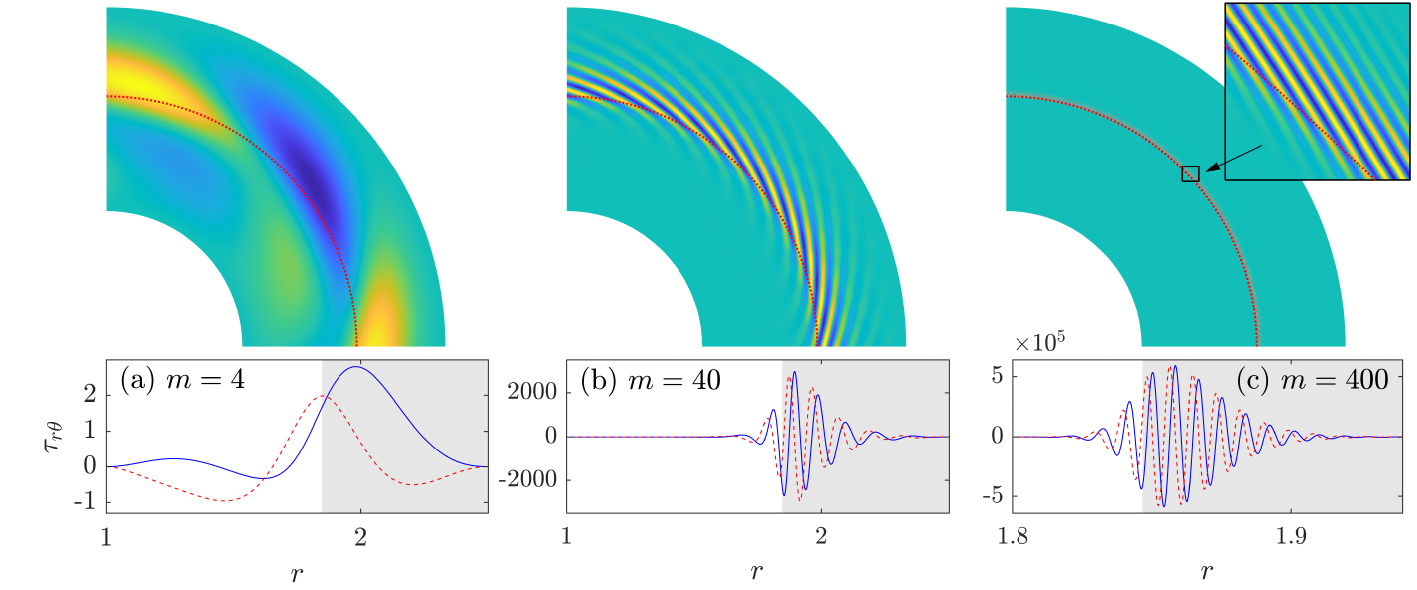}
	\caption{
          Most unstable normal modes for a base state with an outer pseudo-plug. The parameter values are
          $(\Bi,\Wi,\beta,R)=(4,1,\frac12,\frac52)$
          with (a) $m=4$, (b) $m=40$ and (c) $m=400$.
          The upper row shows the spatial pattern of the streamfunction
          of the unstable modes over a quarter of the angular
          domain; in the lower row (panels (a-c)), we show
          the real (solid) and imaginary (dashed)
          parts of $\psi(r)$. The pseudo-plug is shaded light grey
          in the lower row,
          and its border $r=r_p$
          is indicated by dotted lines in the top row.
          For (c), a magnification of the solution is shown
          near $r=r_p$. The modes are normalized so that
        $\int_1^R\psi\;\di r =1$.}
	\label{fig:greig}
\end{figure}
\begin{figure}
  \centering
	\includegraphics[width=\linewidth]{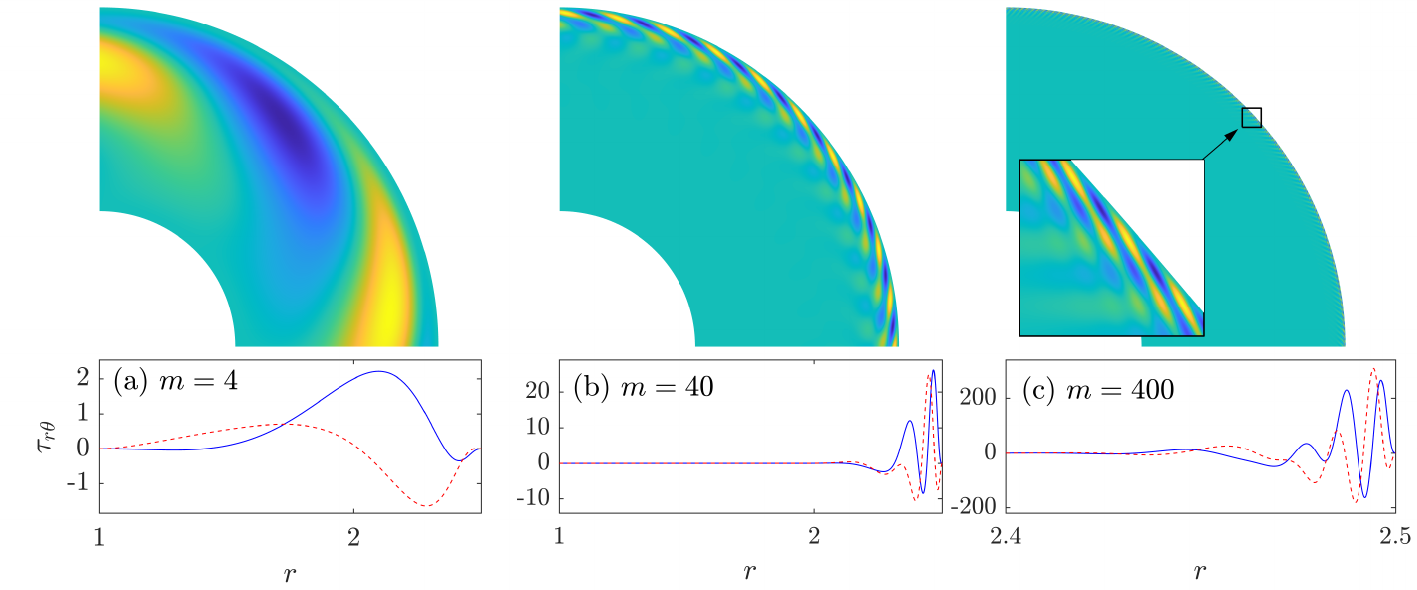}
	\caption{
          A similar set of plots to those in figure \ref{fig:greig}, but for a fully yielded base state. The parameter values are as in figure \ref{fig:greig}, except that $\Bi=0.7$. Again, the wavenumber is (a) $m=4$, (b) $m=40$ and (c) $m=400$.
}
	\label{fig:greigxx}
\end{figure}

The spatial structure of the most unstable modes for $\Bi=4$
and $\Wi=1$ (indicated by the circles in figure \ref{fig:grplot2})
is illustrated in figure \ref{fig:greig}. The streamfunction $\psi(r)$
is displayed
for three values of the wavenumber, taking $\Wi=1$ and the same
remaining parameter values as used for figure \ref{fig:grplot}.
Also shown are renditions of the spatial pattern of the normal modes.
The modes become increasingly localized to the border of the pseudo-plug
for higher wavenumber, so much so that by $m=400$ the eigenfunctions
need to be plotted locally in figure \ref{fig:greig}(c).
The localization of the modes to the border of the pseudo-plug
for high $m$ rationalizes
why growth rates are relatively insensitive to the choice of $A_0$
in figure \ref{fig:grplot2}: 
modes decay spatially well before entering the true plug except when
that feature extends inwards to close to $r=r_p$.

The spatial structure of the normal modes for fully
yielded base states is illustrated in figure \ref{fig:greigxx}.
Again, the modes develop relatively short radial wavelengths
for large $m$. However, the modes now become concentrated near the
outer cylinder, where the shear rate of the base state is smallest.

The short-wavelength character of the instability exposed in
figures \ref{fig:grplot2}-\ref{fig:greigxx} is troubling physically:
without some additional physical effect, one expects perturbations
with a broad range of 
arbitrarily short spatial scales to amplify once a base state
becomes unstable. Though not mathematically ill-posed, this suggests
an excessively sensitive dependence on initial condition and presents
major complications in any numerical simulation of the full nonlinear
dynamics. In \S \ref{sec:nonlin}, we will add polymer stress diffusion
to ease these problems.

\subsection{Short wavelength analysis}\label{sec:LSshort}

The short-wavelength structure of the high$-m$ modes
allows an asymptotic WKB-style
analysis\cite{evpoiy} that provides a prediction for the
high$-m$ growth rate. Essentially, the local dispersion relation
applying at the border of the pseudo-plug
(which is a fourth-order polynomial) must possess
equal pairs of solutions for the local wavenumber, so that
one can construct eigenfunctions that decay to either side.
Some further details of the construction for the current problem
are provided in Appendix \ref{sec:appB}.

\begin{figure}
  \centering
	\includegraphics[width=\linewidth]{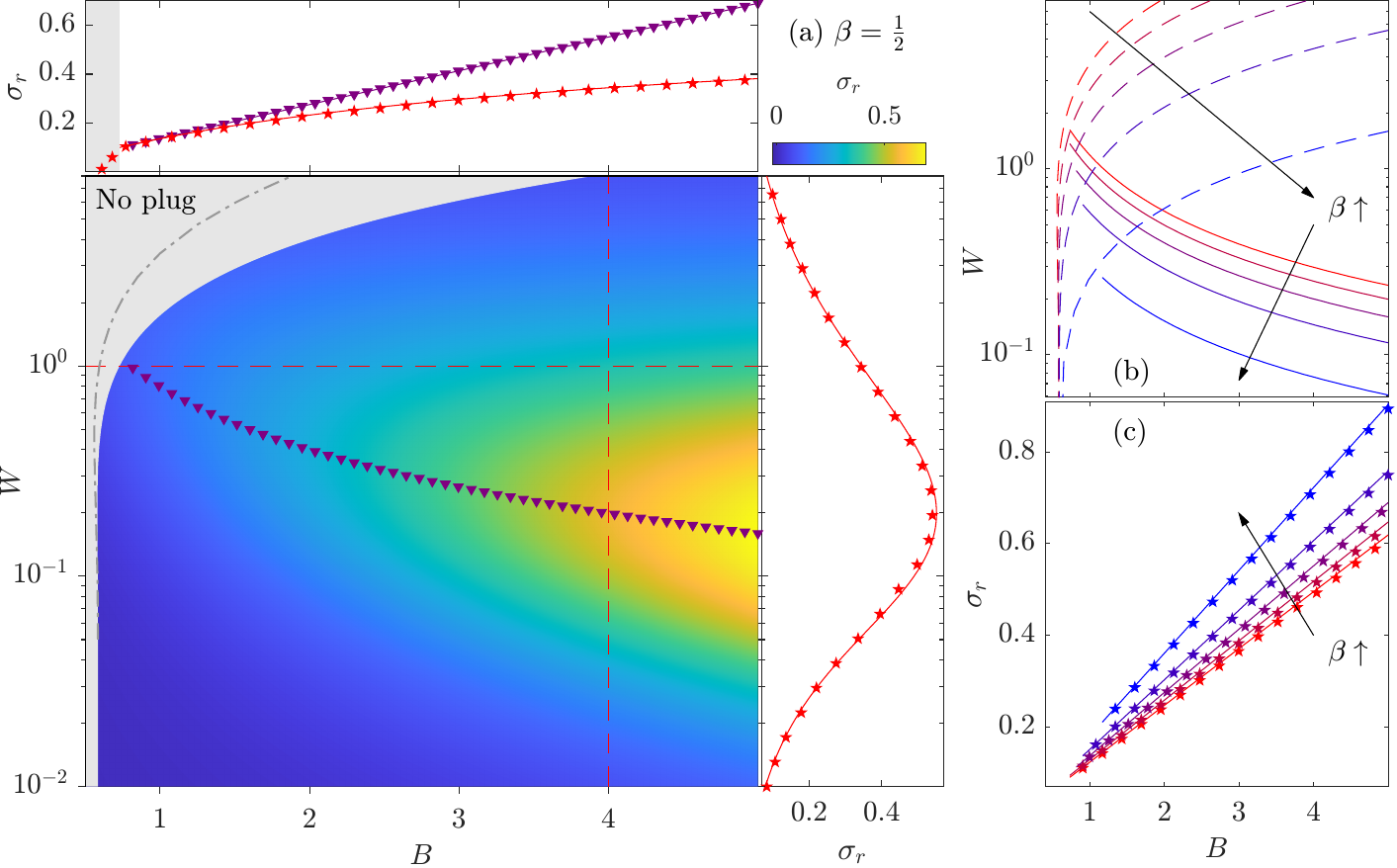}
	\caption{
          (a) Growth rate as a density on the $(\Bi,\Wi)-$plane for
$(\beta,R)=(\frac12,\frac52)$, predicted by short-wavelength asymptotics.
          The light grey shading
          indicates where the plug disappears, and the dot-dashed
          line within it shows the stability boundary computed from
          the full linear stability problem taking $m=500$.
In the smaller panels on the right and above, we plot $\sigma_r$ along the
horizontal and vertical cuts at $\Wi=1$ and $\Bi=4$ (indicated by dashed
lines in the main panel); the stars indicate corresponding numerical
solutions with $m=500$.
For each value of $\Bi$, the maximum growth rate over all $\Wi$
is indicated by the triangles, and plotted in the main and top panels.
The maximum growth rates and the values of $\Wi$ at which they occur
are plotted against $\Bi$
in panels (b) and (c) for $\beta=\{0.1,0.3,0.5,0.7,0.9\}$
(colour-coded from red to blue, and as indicated by the arrows).
The dashed lines in (b) show the corresponding values of $\Wi$
at which the plug disappears (which crudely gauges the position
of the stability boundary).
}
	\label{fig:swtiplo}
\end{figure}

The short-wave analysis leads
to the predictions shown by the stars in
figures \ref{fig:grplot2}(a) and \ref{fig:grplot}(a), and presented
further in figure \ref{fig:swtiplo}. In panel (a) of
the latter, we plot short-wave growth rates over the $(\Bi,\Wi)-$plane
for $(\beta,R)=(\frac12,\frac52)$, together with results
from complementary numerical normal-mode solutions. For each
value of $\Bi$, the growth rates peak at an intermediate value of
$\Wi$; the maximum growth rate and the maximizing Weissenberg number
are shown in figure \ref{fig:swtiplo}(b,c) for different values
of $\beta$. The figure also indicates the value of $\Wi$ for which the
pseudo-plug disappears, at which point the analysis of 
Appendix \ref{sec:appB} no longer applies. 
Growth rates decline to zero shortly above this value, implying 
that it offers a rough guide to the position
of the stability boundary, which is also included in the figure,
computed numerically from the full linear stability theory.

\begin{figure}
  \centering
	\includegraphics[width=\linewidth]{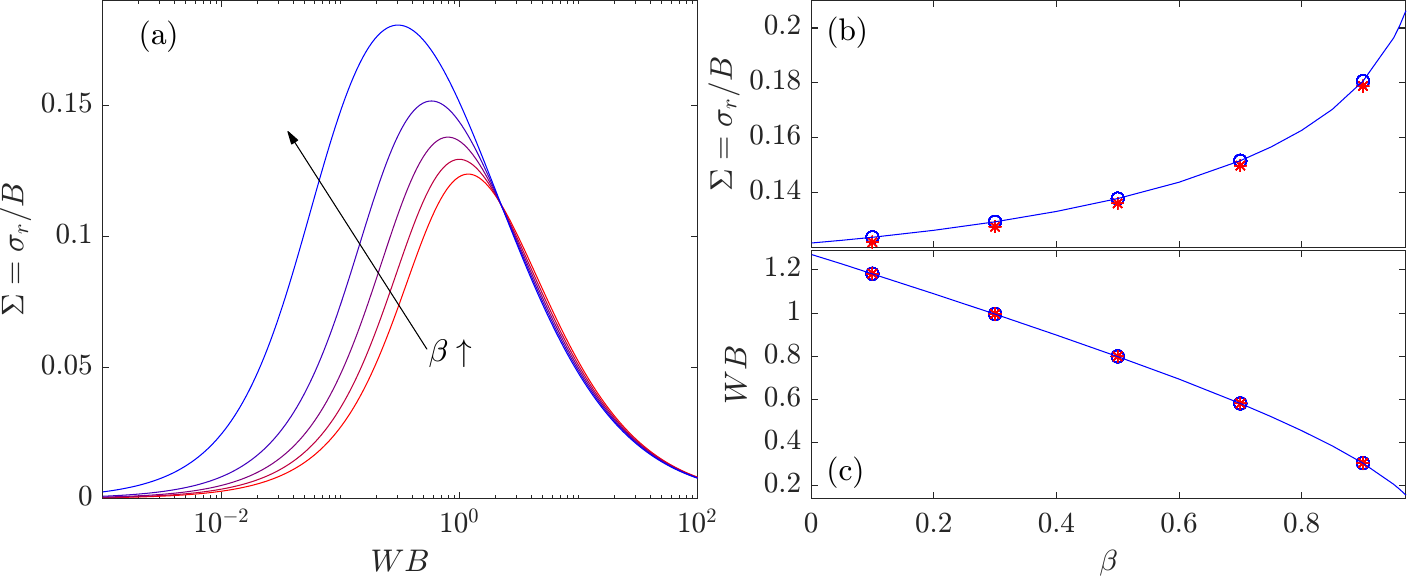}
	\caption{
(a) Scaled growth rate $\Sigma=\sigma_r/\Bi$ against
          $\Wi\Bi$ for $\beta=\{0.1,0.3,0.5,0.7,0.9\}$
          (colour-coded, from red to blue, as indicated by the arrow).
(b,c) The maximum scaled growth rate and the value of $\Wi\Bi$ at which it
occurs, as a function of $\beta$. The data shown by solid lines in
figure \ref{fig:swtiplo}(b,c) are collapsed into the (blue) circles;
the numerical data are indicated by the (red)
stars.
}
	\label{fig:swtiplo2}
\end{figure}

The short-wave analysis actually predicts that the growth rate for $m\gg1$
takes the form
\be
\sigma_r = \Bi \Sigma(\Wi\Bi,\beta),
\ee
with no dependence on $R$, for some function $\Sigma(\Wi\Bi,\beta)$.
This function is plotted for different $\beta$ in figure \ref{fig:swtiplo2}(a),
with the maximum growth rate and the maximizing value of $\Wi\Bi$
plotted in figure \ref{fig:swtiplo2}(b,c). In the latter plots, the data
shown in figure \ref{fig:swtiplo}(b,c) collapse to points.

Note that, although the growth rate becomes independent of $R$ in the
short-wave limit, the conditions dictating when the pseudo-plug
disappears (the dashed lines in figure \ref{fig:swtiplo}(b)) do not,
implying that the stability boundary remains geometry dependent.
One might anticipate that the short-wave growth rate becomes independent
of the size of the annulus in view of the localization of the
modes to $r=r_p$. However, the base state stress and $r_p$ itself
all depend on $R$ through the torque constant $C_\infty$. The fact
that $\Sigma$ does become independent of the annulus geometry
beyond the stability boundary is therefore surprising.

Overall, once the growth rate is scaled by $\Bi$ (implying a
certain characteristic timescale for instability),
we conclude that the linear stability of the base states with
continuous stress profiles is mainly controlled by the parameter combination,
\be
\Wi\Bi \equiv \frac{\lambda\tau_y}{\mu}
= \frac{\mu_p\tau_y}{\mu G} = (1-\beta)\frac{\tau_y}{G},
\ee
where $G=\mu_p/\lambda$ is the elastic modulus.
The ratio $\tau_y/G$ is commonly
referred to as the yield strain in the literature on elasto-viscoplastic
fluids and has been proposed as a key dimensionless
group\cite{moschopoulos2023dynamics,varch,esposito25}. Consequently,
it is perhaps not so surprising that $\Wi\Bi$ controls the linear
stability of elasto-viscoplastic circular Couette flow. Other
than through its dependence in the grouping $\Wi\Bi$, the viscosity ratio $\beta$ plays less of a role,
as illustrated by the relatively mild variations seen with this
parameter in figures \ref{fig:swtiplo} and \ref{fig:swtiplo2}, while the main impact of the geometry of the annulus is ensuring that
the base flow contains a weakly yield region.

\section{Numerical simulations}\label{sec:nonlin}

To explore the nonlinear dynamics prompted by the linear instabilities
detected in \S\ref{sec:LS}, we turn to numerical simulations of
\eqref{eq:gv1}-\eqref{eq:gv9} using the open-source code DEDALUS.
Details of the numerical implementation are provided in Appendix \ref{app:num}.
One key addition in this implementation is the inclusion
of stress diffusion into the constitutive equations
\eqref{eq:comp_rr}-\eqref{eq:comp_tt}. That is, we add
the terms $\kappa\boldsymbol{\nabla\cdot}(\boldsymbol{\nabla \tau})$ to the
right-hand side, where $\kappa$ is a stress diffusion coefficient
and $\boldsymbol{\nabla\cdot}(\boldsymbol{\nabla \tau})$ denotes the tensor Laplacian
expressed in circular polar coordinates (see appendix \ref{app:num}).
Stress diffusion is often included to ease numerical simulations
of viscoelastic fluid flows, and is sometimes justified
by kinetic theory\cite{dubief23,lellep24,lewy26}.
However, the inclusion of such terms requires
additional boundary conditions, whose precise form is not especially clear,
and, awkwardly, can generate new kinds of
instability\cite{beneitez23,beneitez25,lewy26}.

In the present problem, the implementation of the equations in
DEDALUS demands the inclusion of stress diffusion in order to
permit well-resolved, numerically stable computations applying
no-flux conditions at the bounding cylinders. Stress diffusion also
introduces damping at finer spatial scales, which removes the linear
instability that would otherwise arise at short radial
and angular wavelengths. 
However, the nonlinear dynamics may remain sensitive to the value of
$\kappa$; 
we report further on this detail in Appendix \ref{app:diff}.

We also conduct simulations in only part of the annulus in order
improve spatial resolution. That is, we consider periodic
domains with $0\leq\theta<\frac12\pi$. By conducting a few trial
simulations that extended to the full annulus, however, we checked that
the removal of the $m=1$, 2, and 3 modes did not significantly
impact the dynamics.

\subsection{Evolution without an initial polymer stress}

Simulations with EVP fluid models often begin assuming that the
polymer stress vanishes: $\boldsymbol{\tau}(r,\theta,0)=\boldsymbol{0}$.
As pointed out earlier, this assumption is problematic for the construction
of axisymmetrical base states and in testing their linear stability
because stress discontinuities are then expected to develop.
Nevertheless, the relatively weak sensitivity of the growth rates
of linear normal modes towards the detailed plug structure of the
base state (figure \ref{fig:grplot2}(a))
and our earlier results for EVP Poiseuille flow\cite{evpoiy}
suggest that such stress discontinuities will not eliminate
instability. With that in mind, we first conduct simulations
adopting
\be
\tau_{rr}(r,\theta,0) = 0,\quad
\tau_{r\theta}(r,\theta,0) = 0,\quad
\tau_{\theta\theta}(r,\theta,0) = \alpha_0 \sin 4\theta, 
\label{stressIC}
\ee
as initial conditions, where we take $\alpha_0=10^{-4}$,
which introduces a small non-axisymmetrical
perturbation to an otherwise zero polymer stress.
We take the parameter settings
$\{\Wi,\Bi,\beta,R\}=\{1,4,\frac12,\frac52\}$, which,
according to the linear stability results in figure \ref{fig:grplot2},
suggest that a moderately strong instability arises with a growth rate
of order a third. However, because we have not provided
a detailed linear stability analysis in this setting,
the base state being both time-dependent and ultimately discontinuous,
we cannot make any quantitative comparison with normal-mode theory.

\begin{figure}
  \centering
	\includegraphics[width=\linewidth]{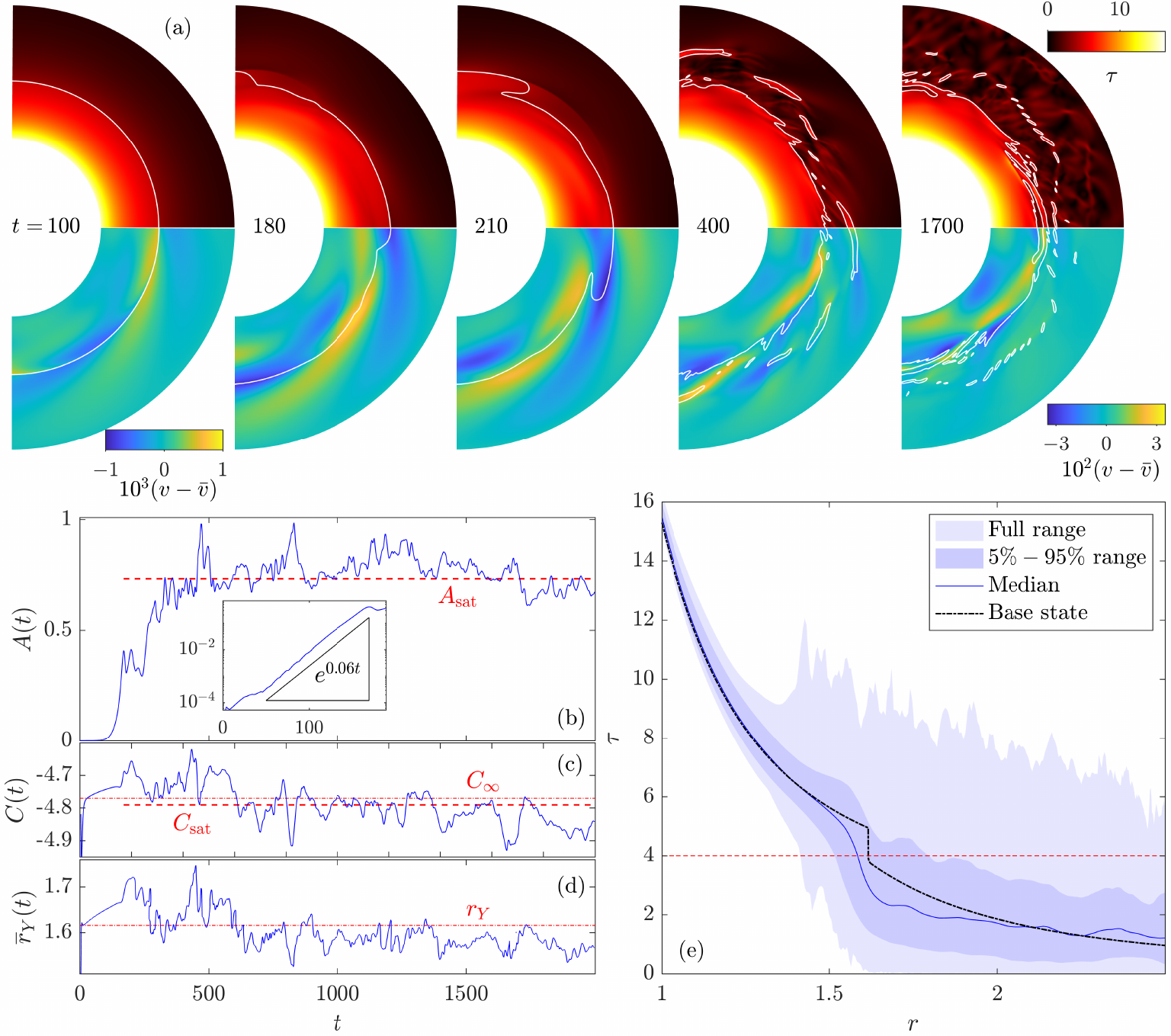}
	\caption{
          Numerical simulation with
          $\{\Wi,\Bi,\beta,R\}=\{1,4,\frac12,\frac52\}$
          and initial stress condition \eqref{stressIC}. Plotted in
          (a) are snapshots of $\tau$ (upper) and $v-\bar{v}$ (lower),
          at the times indicated.
          White contours show the yield surface $\tau=\Bi$.
          The color map for the top row is the same in all panels,
          and indicated by the colorbar on the right. For the
          lower panels, the color map is different for
          the first snapshot, but the same for the others.
          Time series of (b) $A(t)$ in \eqref{eq:Adef}
          (c) $C(t)$, the net torque on the inner cylinder
          in \eqref{eq:Cdef}, and (d) the mean radius of the
          yield surface $\overline{r}_Y(t)$ where $\overline{\tau}=\Bi$.
          In (b), the inset replots the early-time
          data logarithmically and indicates
          exponential growth at rate $0.06$. The dashed lines in (b,c)
          show the saturation levels, obtained from time
          averages over the interval following the first
          peak of $A(t)$; the dot-dashed lines in (c) and (d)
          shows the steady-state torque $C_\infty$
          and yield radius $r_Y$ of the corresponding base state.
          (e) Median of $\tau$ across $0<\theta<\pi/2$ for $1000<t<2000$
          (solid blue), with the darker (blue) shading
          showing the $5^{th}$ and $95^{th}$ percentiles of the
          same $\tau$ data. The lighter (blue) shading
          shows the full range of values of
          $\tau$ over the same time interval. The red dashed line indicates
          $\tau=\Bi$ and the black dot-dashed line shows the discontinous,
          steady-state base solution.
          }
	\label{fig:tpfig}
\end{figure}

Results are shown in figure \ref{fig:tpfig}. The top row of plots
present a sequence of snapshots of the stress invariant $\tau$ and
the non-axisymmetric part of the angular velocity, $v-\bar{v}$,
where the overline denotes angular average:
$$
\overline{(...)} =
\int_0 ^{2\pi} (...) \,\frac{\mathrm{d}\theta}{2\pi}
.
$$
The yield surface (the contour where $\tau=\Bi$) is also indicated.
Below, in panels (b) and (c) we show time series of two convenient
diagnostics: the norm of the non-axisymmetrical
part of $\tau-\overline\tau$
over the whole annulus, 
\be
A(t)
= \left[ \frac{2}{R^2-1} \int_1^R \int_0^{2\pi}
(\tau-\overline\tau)^2 r\di r\frac{\di\theta}{2\pi}\right]^{1/2}
,
\label{eq:Adef}
\ee
and the
net torque on the inner cylinder, defined by
\be
C(t) = \beta r^2\pard{}{r}\left(\frac{\bar{v}}{r}\right) + r^2 \bar\tau_{r\theta}
,
\label{eq:Cdef}
\ee
which is independent of $r$ in view of angular force balance.
The snapshots and temporal diagnostics both reveal the development
of non-axisymmetrical patterns. At early times and after a brief transient,
non-axisymmetrical perturbations grow approximately exponentially,
as indicated by the amplitude measure $A(t)$.
The observed  growth rate is close to
$0.06$, which is somewhat smaller than the
spread for $m=4$ seen in figure \ref{fig:grplot2}(a). A discrepancy with the growth rate from the normal-mode theory is to be expected given the different form of base state, as well as the fact that the base state is evolving concurrently with the perturbation in simulations. 

Exponential growth saturates for times of order a hundred, at which
stage a spatio-temporally complicated state emerges
in which wave-like perturbations continually develop in the vicinity
of the yield surface, breaking down into
structures with shorter spatial scales that migrate into the outer plugged
regions. Averaging the time series of $A(t)$ and $C(t)$
over the interval following the first maximum of the amplitude measure
leads to two diagnostics of the saturated state, $A_{\rm sat}$ and $C_{\rm sat}$,
as illustrated in figure \ref{fig:tpfig}(b,c). The latter is slightly
smaller than the steady-state torque experienced for the axisymmetric
base state (also shown).

The last panel (e) of figure \ref{fig:tpfig}
shows the radial profile of
later-time fluctuations of the stress invariant: the solid line
shows the median of $\tau$ over $0<\theta<\frac12\pi$ and $t>10^3$,
an interval long enough to ensure that $A(t)$ has reached
what looks to be a statistically steady state.
The lighter shading indicated the full variability of $\tau$
over the same range of time and angle; the stress invariant clearly
fluctuates from relatively low levels to values well above the
yield stress. However, the extrema in stress that are reached are
relatively rare across the full annulus and time window,
as can be seen from the darker shading, which shows
where $\tau$ fluctuates from the $5^{th}$ to the $95^{th}$ percentile, and occupies a much tighter band around that median.
Evidently, fluctuations are strongest near the yield surface
and over the inner part of the plug, becoming weaker near the
bounding cylinders.

\begin{figure*}
  \centering
	\includegraphics[width=\linewidth]{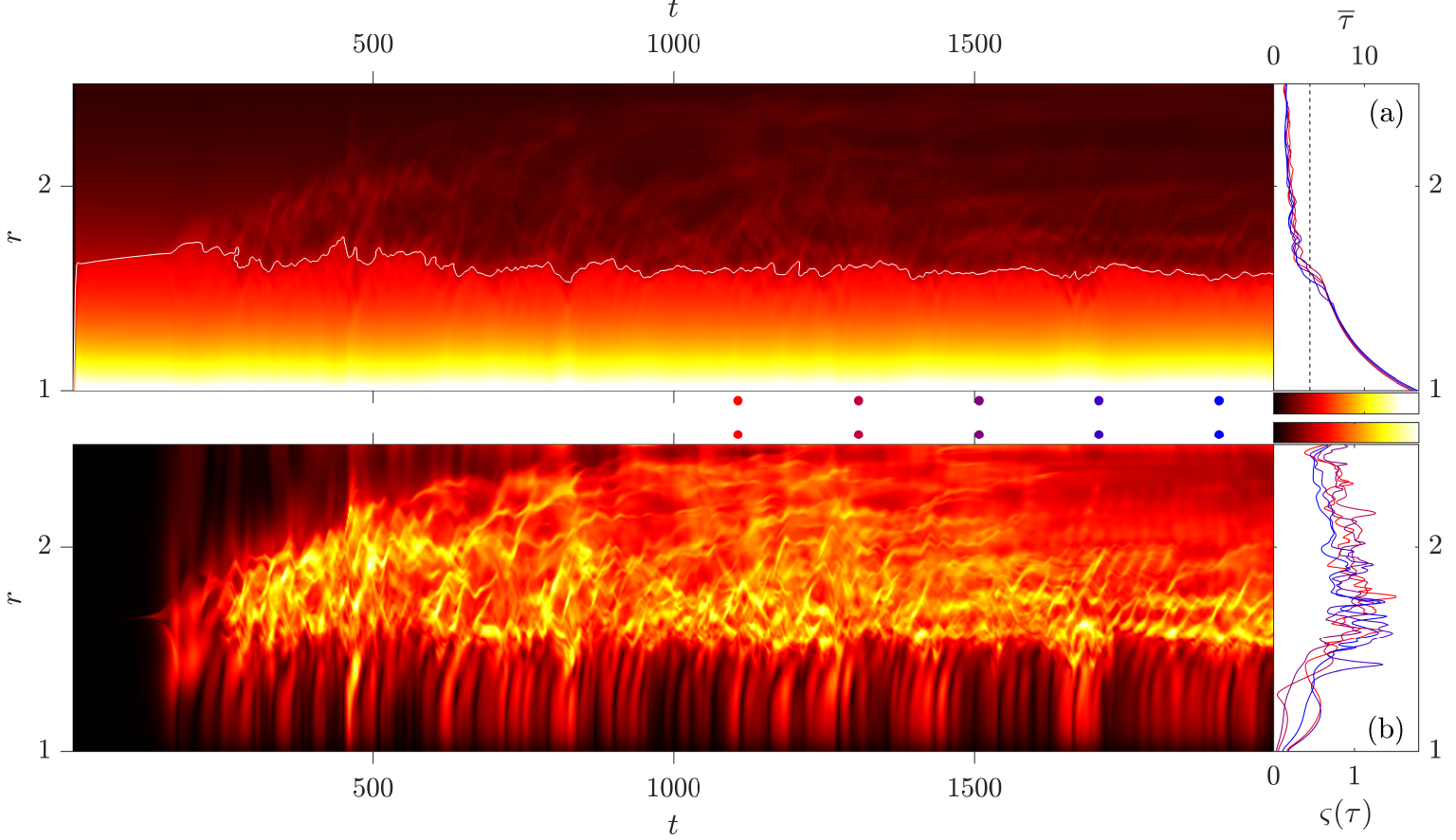}
	\caption{
          Plots of (a) $\overline\tau$ and (b) $\varsigma(\tau)$,
          the standard deviation of $\tau$ over angle,
          as densities over the $(t,r)-$plane,
          for the simulation in figure \ref{fig:tpfig}. For each
          panel, we plot snapshots on the right at the times
          indicated by the dots (colour-coded, from red to blue).
          }
	\label{fig:xtpfig}
\end{figure*}

Further details of the mean stress and its fluctuations are shown
in figure \ref{fig:xtpfig}. Panel (a) displays the angular average
$\overline\tau$ on a space-time diagram. Over the duration
of the nonlinear phase ($t>250$ or so), the mean position
of the yield radius $r=\overline{r}_Y(t)$, as given by $\overline{\tau}=\Bi$,
migrates slightly inwards, as shown more clearly in
figure \ref{fig:tpfig}(d). This position appears closely connected
with the torque $C(t)$ (compare panels (c) and (d) in figure \ref{fig:tpfig}).

Figure \ref{fig:xtpfig}(b) shows instead
$\varsigma(\tau)$, the standard deviation
of $\tau(r,\theta,t)$ over angle, as another density plot.
In addition to the concentration of the fluctuations
towards the yield surface, this plot also reveals two
other notable features: first, the fluctuations appear
to be episodic over the yielded part of the annulus.
Second, there is a clear signature of propagation
for non-axisymmetrical disturbances (tilted streaks of
enhanced values of $\varsigma(\tau)$) over shorter
distances within the plug. These propagating streaks are
symptomatic of the fingers of yielded fluid that grow from the yield
surface and protrude into the plug.

\subsection{Evolution from a steady base state with an outer
pseudo-plug}

\begin{figure*}
  \centering
	\includegraphics[width=\linewidth]{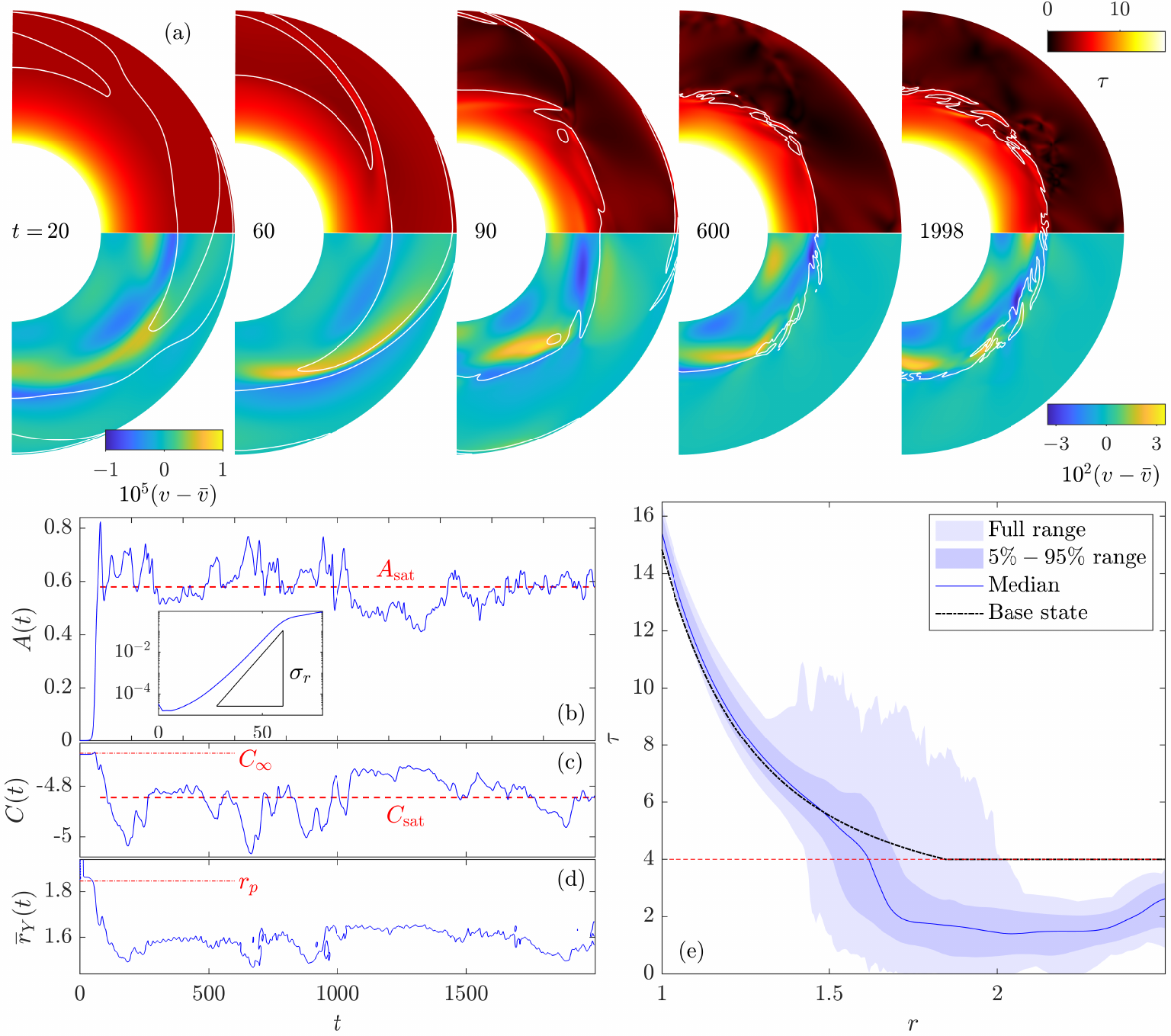}
	\caption{A similar set of plots to those in figure \ref{fig:tpfig},
          but for a simulation in which the solution is initialised using
          a steady base state with an outer pseudo-plug; the initial profile
          of $\bar{\tau}$ is shown in (e) by the black dot-dashed line,
          the border of the pseudo-plug $r_p$ is indicated by the
          dot-dashed line in (d).
          The inset to (b) shows a logarithmic plot at early times and
          the predicted growth rate from linear theory, $\sigma_r\approx0.26$. 
          }
	\label{fig:ppfig}
\end{figure*}

A potentially clearer comparison of linear theory with
the early-time dynamics in a numerical simulation
is shown in figure \ref{fig:ppfig}. In this example, the
numerical simulation begins from a weakly perturbed,
steady base state solution constructed using the same parameters
as figure \ref{fig:tpfig}. The base state has an outer pseudo-plug
(as would be generated by a pre-shear with $A_0\gtrsim4.42$),
and the simulations begin with
\be
\tau_{rr}(r,\theta,0) = 0,\quad
\tau_{r\theta}(r,\theta,0) = T_{r\theta}(r),\quad
\tau_{\theta\theta}(r,\theta,0) = T_{\theta\theta}(r)+\alpha_0 \sin 4\theta,
\label{stressIC2}
\ee
where $T_{r\theta}(r)$ and $T_{\theta\theta}(r)$ denote the non-trivial,
steady, base-state stress components, and again we take to be $\alpha_0=10^{-4}$. The base state and its linear
stability characteristics correspond to the circles shown in
figure \ref{fig:grplot2}, and the normal-mode eigenfunctions for
particular angular wavenumbers are displayed in figure \ref{fig:greig}.

The non-axisymmetrical initial perturbations again seed the exponential
growth of unstable modes; the observed growth rate in the simulations
is now much closer to that expected from linear theory (see panel (b)).
Awkwardly, however,
the linear theory assumes that the pseudo-plug of the base state
is everywhere slightly above the yield stress. By contrast,
the initial stress conditions in \eqref{stressIC2} immediately lead to
a fraction of the outer part of the annulus plugging up. Consequently,
the early-time exponential growth observed in the simulation
becomes slightly offset from
the predictions of linear theory, even after initial transients subside.
Regardless, the linear instability again saturates to leave a
similar spatio-temporally complicated state as that observed
with stress-free initial conditions.

In finer detail, the two simulations in figures
\ref{fig:tpfig} and \ref{fig:ppfig} differ in their saturated nonlinear
state: in the former, the stress fluctuations saturate at higher amplitude and
penetrate further into the outer plug,
whereas a wider sector of the annulus at the outer cylinder
becomes fully plugged in the latter.
Evidently, the growth of the linear instability is
not sufficient to erase the memory of the stress history over the
plugged regions. 

\begin{figure*}
  \centering
	\includegraphics[width=\linewidth]{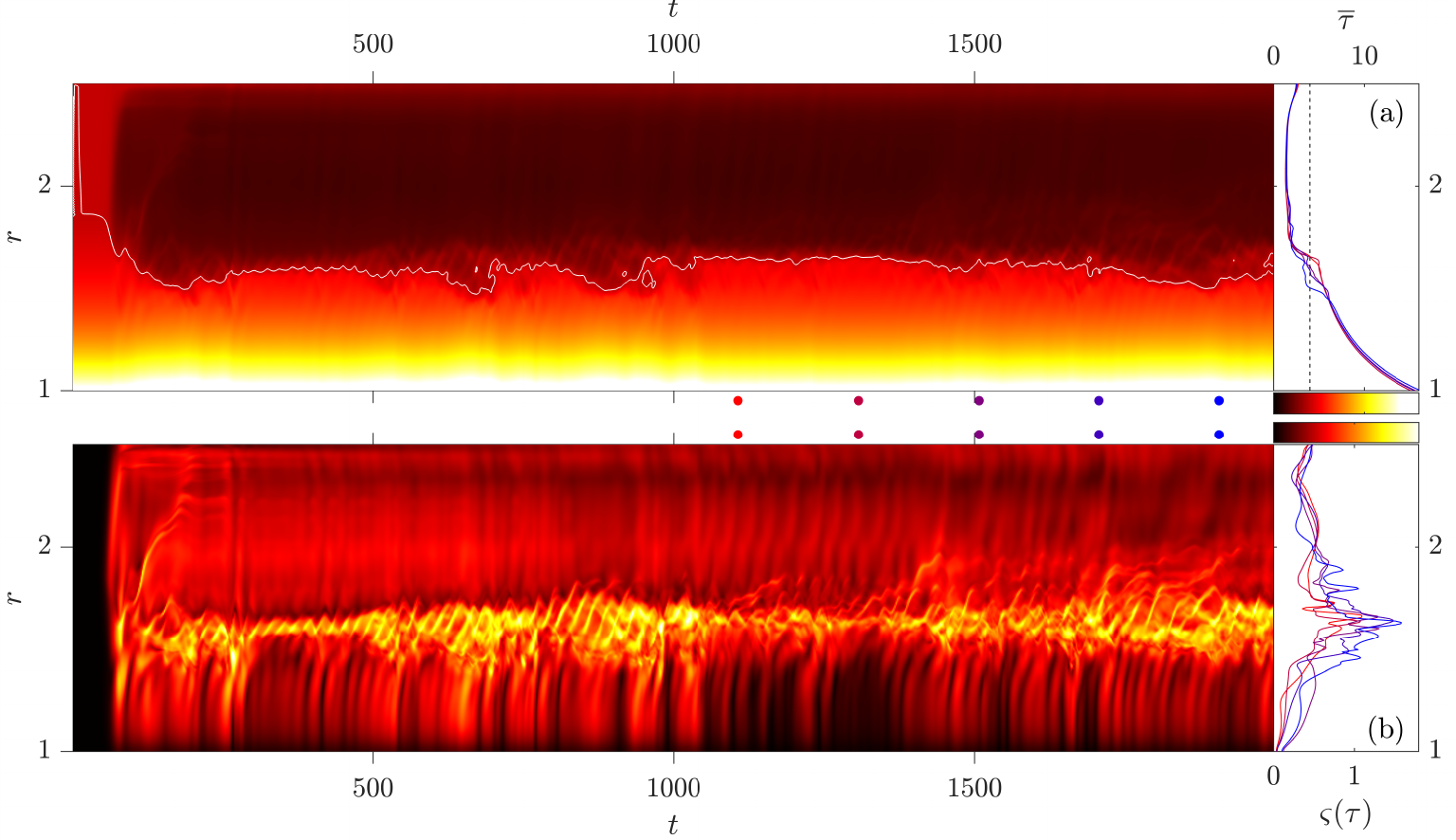}
	\caption{
          A similar pair of plots to those in figure \ref{fig:xtpfig},
          but for the simulation (figure \ref{fig:ppfig}) which was initialised from a base state with an outer pseudo-plug.
          }
	\label{fig:xppfig}
\end{figure*}

Figure \ref{fig:xppfig} repeats the plots in
figure \ref{fig:xtpfig} for the simulation starting with a pseudo-plug.
In this case, the mean yield surface shows little notable
radial migration, in line with the torque time series
(see figure \ref{fig:ppfig}(c,d)). 
The saturated torque $C_{\rm sat}$ is also rather lower
than the steady-state torque of the base state $C_\infty$ in this
example, and the mean yield radius $\overline{r}_Y(t)$
migrates much further in than the edge of the original pseudo-plug.
The plot of the angular fluctuations shown in figure \ref{fig:xppfig}
again highlights the concentration to the yield surface as well
as the episodic variability over the yielded region and propagating
disturbances over the plug. The fingers of yielded fluid now
protrude less far into the plug.

\subsection{Evolution of a fully yielded annulus}

Finally, we present a third sample simulation in which the initial
state commences from a weakly perturbed, steady, fully yielded base state.
This example has $\Bi=0.7$ and corresponds to the linear stability analysis
denoted by triangles in figure \ref{fig:grplot2}(c), with the sample
eigenmodes shown in figure \ref{fig:greigxx}.
The initial stress conditions are again given by \eqref{stressIC2},
but for this new fully yielded base state.

\begin{figure*}
  \centering
	\includegraphics[width=\linewidth]{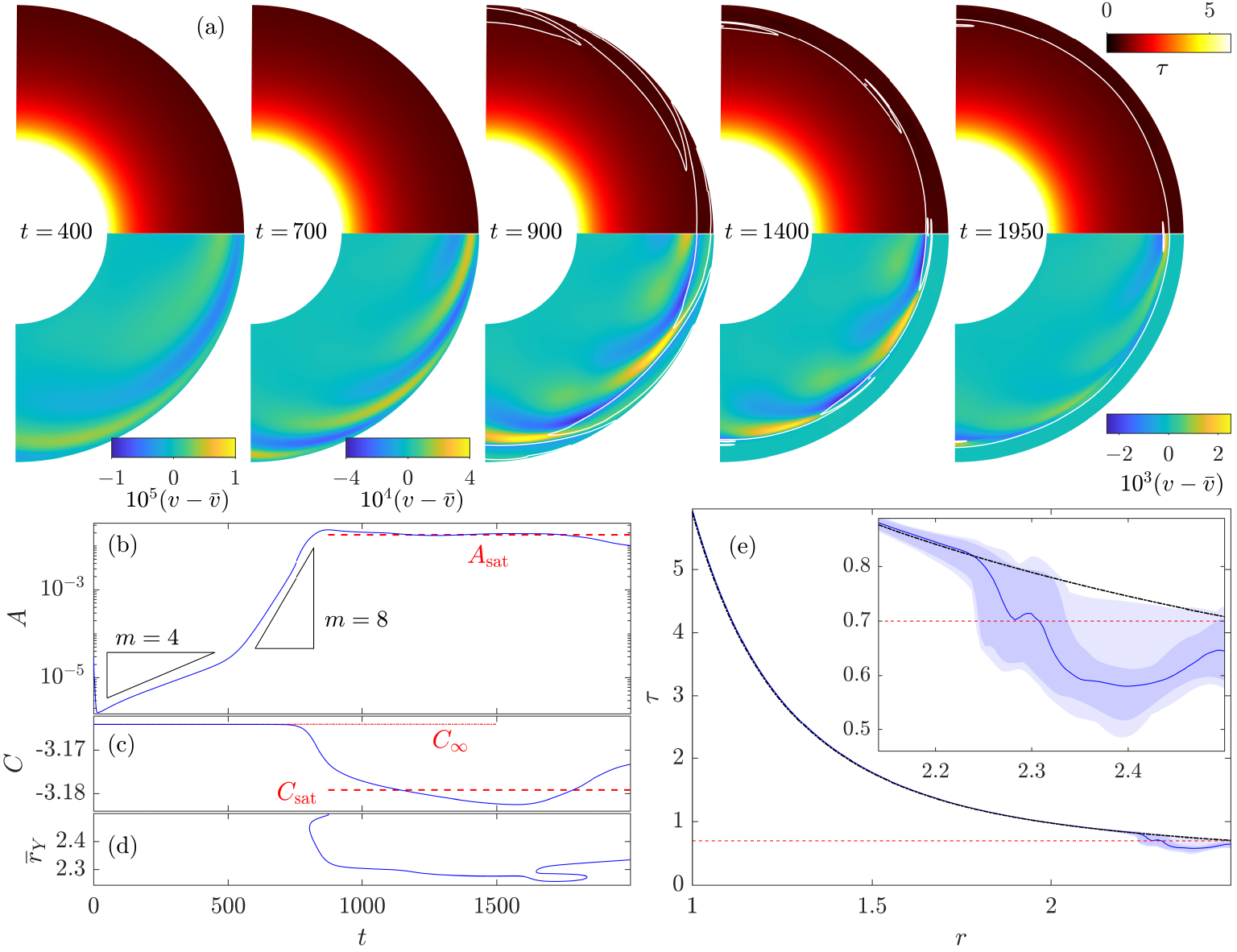}
	\caption{
          A similar set of plots to those in figure \ref{fig:tpfig},
          but for a simulation in which the solution is initialised using
          a steady, fully yielded base state with $\Bi=0.7$;
          the initial profile
          of $\bar{\tau}$ is shown in (e) by the black dot-dashed line.
          In (b), 
          the predicted growth rates of the normal modes with
          $m=4$ ($\sigma_r\approx0.006$) and $m=8$ ($\sigma_r\approx0.024$) are indicated.
          For (e), the inset shows a magnification near the outer
          cylinder.
          For the first two snapshots of $v-\bar{v}$, the colormap
          is different, and indicated by the colorbars shown at the
          bottom of each panel. 
          }
	\label{fig:npfig}
\end{figure*}

As emphasized in figure \ref{fig:grplot2}(c), the low wavenumbers
for this base state are only relatively weakly unstable. Indeed,
although the simulation is initialized with perturbations with $m=4$,
these modes do not dominate the linear phase of evolution: after a brief
period over which the growth in $A$ follows the expected
linear growth for $m=4$ (for $t<450$,
see figure \ref{fig:npfig}(b)), the mode with
$m=8$ takes over to accelerate the exponential amplification
of the non-axisymmetrical perturbations and convey then
into the nonlinear regime.
Simultaneously, the
spatial pattern of $v-\bar{v}$ visibly doubles its angular wavenumber
(compare the first two
snapshots in figure \ref{fig:npfig}(a)).

The saturation of the linear instability in this sample simulation
leads to a much less complicated nonlinear state than seen in
figures \ref{fig:tpfig} and \ref{fig:ppfig}: although the stress
become suppressed near the outer cylinder to create a long-lived plug
there, the fluctuations near the resulting yield surface are
much weaker, with the bulk of the yielded part of the annulus remaining
axisymmetrical and steady.
Evidently, although the linear instability extends to base
states without plugs and drives persistent non-axisymmetrical
patterns, the impact on the flow dynamics is much less significant
than when the corresponding base state contains a plug or pseudo-plug.

\subsection{Varying $\Bi$}\label{sec:bvar}

\begin{figure}
  \centering
	\includegraphics[width=.9\linewidth]{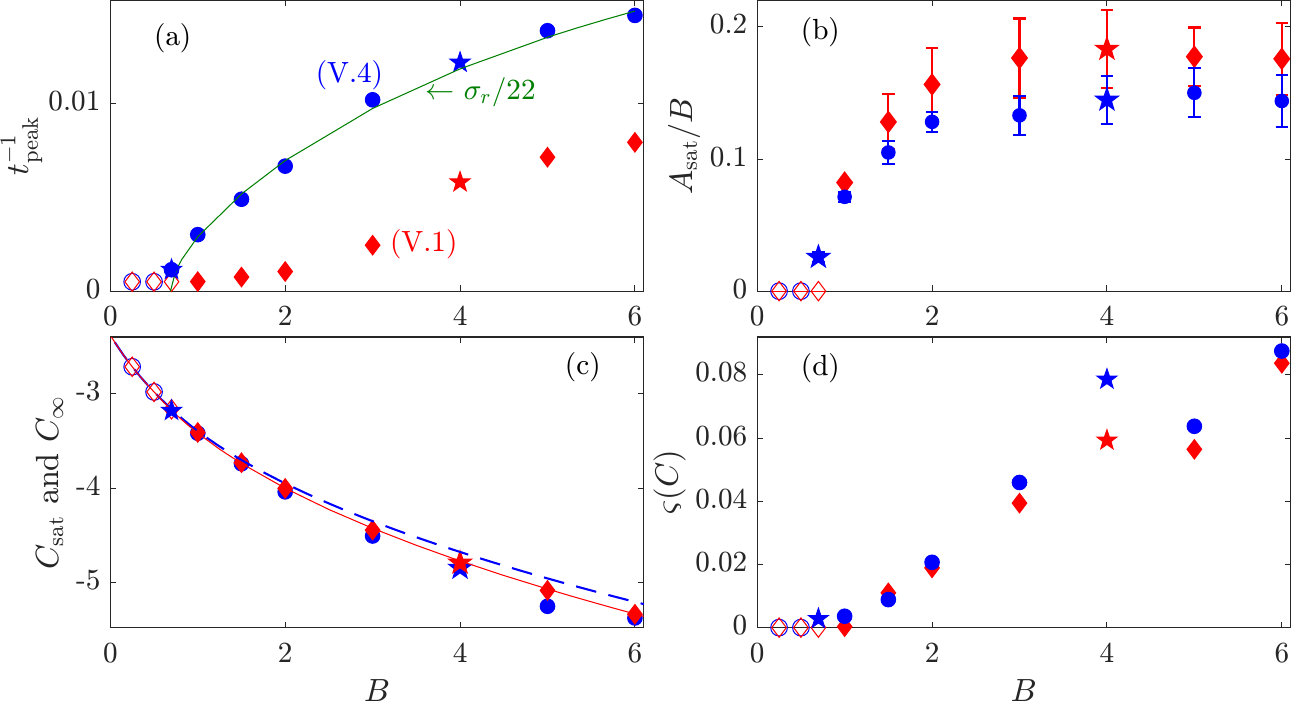}
	\caption{Data from two suites of simulations with varying $\Bi$,
          for $(\Wi,\beta,R,\kappa)=(1,0.5,2.5,10^{-6})$.
          In one suite (red diamonds), the initial conditions
          are given by \eqref{stressIC}; for the blue circles,
          the initial condition is given by \eqref{stressIC2}.
          (a) $t_{\mathrm{peak}}^{-1}$, where $t=t_{\mathrm{peak}}$ is the time at
          which $A$ has its first maximum (discarding
          any maxima below $0.01$ to remove initial transients).
          (b) $A_{\rm sat}$, the mean value of $A$ across the interval
          $t_{\mathrm{peak}}<t<2000$, with error bars indicating one standard
          deviation from the mean. (c,d) Mean $C_{\rm sat}$
          and standard deviation $\varsigma(C)$ of the torque
          across the same time interval. The stars shows the
          simulation from figures \ref{fig:tpfig}-\ref{fig:npfig}.
          In (a), the solid line shows the computed linear growth rate
          for $m=4$ from figure \ref{fig:grplot2}(c), scaled by 22.
          The solid and dashed lines in (c) show the steady-state
          torques $C_\infty$ from figure \ref{fig:regiplot}.
          }
	\label{fig:Bvar}
\end{figure}

For a broader view of the complex nonlinear dynamics beyond
the three preceding illustrative examples, we present results
from a suite of simulations in which we vary $\Bi$,
holding fixed the other parameters. 
This suite follows the same path in parameter space as the base-state
sweep in figure \ref{fig:regiplot} and 
the linear stability survey in figure \ref{fig:grplot2}.
The results are shown in figure \ref{fig:Bvar}. Two suites of
simulations are presented, computed from a
weakly perturbed base state with either zero polymer stress initial
conditions (equation \eqref{stressIC}),
or a perturbed non-diffusive steady state (equation \eqref{stressIC2}).

Figure \ref{fig:Bvar}(a) shows $t_{\mathrm{peak}}^{-1}$, the inverse of the time taken for $A$ to reach a peak. This would be proportional to the growth rate were the growth to be purely exponential up until $t=t_{\mathrm{peak}}$; whilst the initial phase is never perfectly exponential, figure \ref{fig:Bvar}(a) still provides an indicative measure of the impact of $B$ on the early-time growth rate. In line with the growth rates from linear theory (as given in figure \ref{fig:grplot2}(c), and reproduced in figure \ref{fig:Bvar}(a) with a scaling), $t_{\mathrm{peak}}^{-1}$ increases monotonically as $B$ is increased. For $B\leq0.5$, no unstable growth is observed, and no peak in $A$ is recorded. This is also consistent with
linear theory (figure \ref{fig:grplot2}), where the critical
Bingham number for instability with these parameters 
was found to be $\Bi\approx0.6$. The approximate growth rates are systematically lower for the suite of simulations using the initial conditions \eqref{stressIC} compared with those using the pre-stressed initial conditions \eqref{stressIC2}, but the general trends are similar. 

Figure \ref{fig:Bvar}(b) illustrates how the saturation amplitude $A_{\rm sat}$,
varies with $\Bi$, defining this level
as the average of $A(t)$ over $t_\mathrm{peak}<t<2000$. There is a steady
increase in the mean and standard deviation of
$A$ as $\Bi$ is increased.
Scaling both with $\Bi$, however, removes most of this trend, as shown
in the figure (where the error bars indicate the standard deviations).
The mean magnitude and standard deviation of the torque $C$
mostly increase with $B$, with the means lying close to
the steady-state values $C_\infty$ for base states
computed with zero initial polymer stress (see figure \ref{fig:Bvar}(c,d)).

For values of $\Bi$ close to the critical value for instability, the
complex state that emerges possesses a relatively low stress
fluctuation level. Indeed, $A_{\rm sat}$ appears to increase continuously
from onset, suggesting a supercritical transition.
The difference between the data for the two suites of simulations in figure \ref{fig:Bvar}(b,c) is relatively small. The slightly lower saturation level
in simulations with pre-stressed initial conditions may reflect the
relatively lower level of stress fluctuations over the near-wall plug
({\it cf.} figures \ref{fig:xtpfig} and \ref{fig:xppfig}). However, at
each $\Bi$, $A_{sat}$ for the two simulations is within a standard deviation,
implying that the detailed dynamical differences are not strongly reflected
in global measures of the stress fluctuation amplitude.

\section{Discussion}

We have explored the linear stability of
two-dimensional circular Couette flow of an elasto-viscoplastic
at zero Reynolds number. For this task, we first explored
axisymmetrical base states, which are non-unique and can possess
discontinuous stress profiles when fluid falls below the yield
stress somewhere within the annulus\cite{cheddadi08,cheddadi12}.
For linear stability analysis, we considered only a subset of this
wide array of equilibria because discontinuous base states present
significant challenges in linear theory\cite{evpoiy}.
In particular,
we focussed on base states for which fluid near the outer wall flows
as a marginally yielded `pseudo-plug'. Such base states can be generated by
briefly raising the cylinder rotation rate to pre-shear the fluid, and
possess continuous stress profiles. Linear stability analysis
then follows more straightforwardly.

Base states with
pseudo-plugs arise for sufficiently large yield stresses and
are found to be unstable at order-one Weissenberg numbers.
We also found that fully yielded base flows with a low-shear region
near the outer wall can be unstable. That said, growth rates
are highest when there is a sizeable pseudo-plug.
Concerningly, and as
found in Poiseuille flow\cite{evpoiy}, the growth rate of this
elasto-viscoplastic instability is maximized at the highest streamwise
wavenumbers. The most unstable modes are
spatially localized to the boundary of
the pseudo-plug, or to the outer wall when the base
state is fully yielded. Whilst a simple mechanism for
such instabilities remains to be identified
under general physical conditions\cite{evpoiy},
the localization is a distinctive feature,
common to both channel and circular Couette flow. 

To explore the nonlinear dynamics of the instability, we performed
two-dimensional numerical simulations, including a
small level of polymer stress diffusion. In addition to
easing computations, such stress diffusion 
removes instability at the shortest spatial scales, permitting
well-resolved numerical solutions.
The simulations also allowed us to advance beyond steady
base states with pseudo-plugs and
consider the stability of more general axisymmetrical configurations
evolving towards profiles that would become discontinuous without
stress diffusion.
In particular, we explored the stability of base states
evolving from zero initial polymer stress, which is
a commonly used initial condition in elasto-viscoplastic
flow modelling. Both kinds of base states proved to be linearly unstable.

For base states with either true plugs or pseudo-plugs,
the linear instability becomes saturated by nonlinearity,
triggering spatio-temporally complex flow states
composed of a strongly yielded region at the inner wall
and an unyielded plug tethered to the static outer wall;
stress fluctuations are continually generated
at the intervening yield surface, acting as the motor driving complexity.
By comparison, fully yielded base states suffer a much milder instability
and the nonlinear dynamics are notably less complicated, although
a thin plug still forms against the outer wall.
Thus, whereas the presence of a plug or pseudo-plug in the base state
is not essential to trigger linear instability,
an extensive unyielded region is important for driving complex dynamics.

In simulations of elasto-viscoplastic Poiseuille flow \cite{evturbo},
we found that the details of the
complex nonlinear states were relatively insensitive
to the initial conditions. In particular,
we found that the dynamics was insensitive to
whether it developed from zero initial polymer stress
or a steady base state with a pseudo-plug. By contrast,
here we have found that the initial conditions used for the
underlying base state do play a more significant role, 
with a zero initial polymer stress promoting slightly
more complex dynamics. We attribute this difference to the plug:
in circular Couette flow,
the plug is relatively wide and locked to the stationary
outer wall, enhancing the effective memory of the stress history.

Our main effort here has been to explore the impact of the yield stress 
in concert with elasticity, leading us to focus on the effect of varying the Bingham number when conducting numerical simulations. We provided a more comprehensive exploration of other parameters in our linear stability analysis. The most significant
effect of varying annular geometry (the outer
cylinder radius $R$) is to limit when low shear-rate regions
appear in the base flow, thereby
impacting linear instability. Beyond that effect, our normal-mode analysis
establishes that the timescale for instability is set by
$\mu/\tau_y$ and controlled otherwise
by the product, $\Wi\Bi=(1-\beta)\tau_y/G$,
where $\mu=\mu_p+\mu_s$ is the total viscosity,
$\beta=\mu_s/\mu$ is the viscosity ratio and
$\tau_y/G$ is the yield strain (the ratio of
the yield stress $\tau_y$ and elastic modulus $G$).
Otherwise, both $\beta$ and $R$ play relatively minor roles in the normal-mode theory, leading
us to adopt representative values, $\{\beta,R\}=\{\frac12,\frac52\}$, for these parameters throughout
our study. Nevertheless, in view of the uncertainty in the former for real
fluids\cite{agrawal25}, and that smaller values are often employed (\textit{e.g.}
Refs.~\onlinecite{franca2024,mousavi24}), we have run trial numerical simulations adopting values over the range $0.1\leq \beta\leq 0.5$ without observing a significant change to the dynamics (see Appendix \ref{app:beta}). 

We close by emphasizing that
our current and previous results\cite{evpoiy,evturbo}
demonstrate that complex dynamics arises
through linear instability in inertialess flows
of elasto-viscoplastic fluids described by Saramito's first model (the `Saramito-Bingham' model). 
What remains unclear is
whether the dynamics is physical or an artefact of the constitutive model.
Indeed, the short-wave character of the instability is problematic.
Experiments with real yield-stress fluids may shed light on this issue,
although extensions of the theory to three dimensions would then be demanded.
Awkwardly, other viscoelastic-type flow instabilities
can appear in Taylor-Couette flow\cite{larson90}
or be generated by polymer stress diffusion\cite{beneitez23}
to confound matters further. Regardless, computations of elasto-viscoplastic
flows using this version of Saramito's model are prone to the development
of excessively short spatial scales even at low Reynolds number, which
likely adds to the complications of performing numerical simulations.
Other elasto-viscoplastic constitutive
laws may well also suffer a similar fate (\textit{e.g.}
Refs.~\onlinecite{saramito09,saraFENEP,KDR}), but this remains to be seen.

\acknowledgements
JDS has been supported by a Leverhulme Trust Study Abroad Studentship.
NJB acknowledges NSERC Discovery Grant RGPIN-2022-03705.
We attempted to use AI to help with the analysis and preparation
of this manuscript, but largely failed miserably; we did manage to
get Claude (Anthropic) to translate some equations into LATEX for the document
and adapt codes for the base states and linear stability to the
present problem. But it wasn't pretty. Claude did help in deriving the
formulae in Appendix \ref{sec:appB}.

\begin{appendix}

\section{Numerical details}\label{app:num}


To ease numerical simulations using the pseudo-spectral solver DEDALUS, we introduce polymer stress diffusion to the constitutive equations. 
Specifically, we add $\kappa (\nabla^2\boldsymbol\tau)_{ij}$ 
to \eqref{eq:comp_rr}-\eqref{eq:comp_tt}, corresponding to
the $(i.j)$-components of the tensor Laplacian of $\boldsymbol\tau$
with diffusion coefficient $\kappa$.
Explicitly, these components are given by
\begin{align}
(\nabla^2\boldsymbol\tau)_{rr}
&= \dfrac1r\dfrac{\partial}{\partial r}\!\left(r\dfrac{\partial \tau_{rr}}{\partial r}\right)
+\dfrac{1}{r^2}\dfrac{\partial^2 \tau_{rr}}{\partial\theta^2}-\frac{4}{r^2}\frac{\partial\tau_{r\theta}}{\partial\theta}
-\frac{2}{r^2}(\tau_{rr}-\tau_{\theta\theta}),
\label{eq:lap_rr}\\
(\nabla^2\boldsymbol\tau)_{\theta\theta}
&= \dfrac1r\dfrac{\partial}{\partial r}\!\left(r\dfrac{\partial \tau_{\theta\theta}}{\partial r}\right)
+\dfrac{1}{r^2}\dfrac{\partial^2 \tau_{\theta\theta}}{\partial\theta^2}+\frac{4}{r^2}\frac{\partial\tau_{r\theta}}{\partial\theta}
+\frac{2}{r^2}(\tau_{rr}-\tau_{\theta\theta}),
\label{eq:lap_tt}\\
(\nabla^2\boldsymbol\tau)_{r\theta}
&= \dfrac1r\dfrac{\partial}{\partial r}\!\left(r\dfrac{\partial \tau_{r\theta}}{\partial r}\right)
+\dfrac{1}{r^2}\dfrac{\partial^2 \tau_{r\theta}}{\partial\theta^2}+\frac{2}{r^2}\frac{\partial(\tau_{rr}-\tau_{\theta\theta})}{\partial\theta}
-\frac{4}{r^2}\tau_{r\theta}.
\label{eq:lap_rt}
\end{align}
In each of \eqref{eq:lap_rr}-\eqref{eq:lap_rt}, the first two terms equal the scalar Laplacian of the corresponding stress component, and the remaining terms arise from differentiation of the basis vectors.
Additional stress boundary conditions are required at the inner and outer walls; we enforce
\begin{equation}
	\frac{\partial\tau_{rr}}{\partial r} = \frac{\partial\tau_{r\theta}}{\partial r} = \frac{\partial\tau_{\theta\theta}}{\partial r} = 0 \quad \mbox{at}\quad r = 1,R. 
\end{equation}

We also smooth the switch at the yield stress in the constitutive law
to enable computation in DEDALUS, replacing $f$
in \eqref{eq:comp_rr}-\eqref{eq:comp_tt}
with 
\begin{equation}
	\tilde{f} = \frac{1}{2}\left[1+\tanh\left(\frac{\tau-\Bi}{\epsilon}\right)\right]\left(1-\frac{\Bi}{\tau}\right),\label{eq:fdefsmooth}
\end{equation}
where we use $\epsilon=10^{-4}$, which is chosen to be small enough that $\tilde{f}$ closely approximates $f$. We adopted the same approach previously in DEDALUS simulations of Poiseuille flow\cite{evturbo}. 

\begin{figure}
  \centering
	\includegraphics[width=\linewidth]{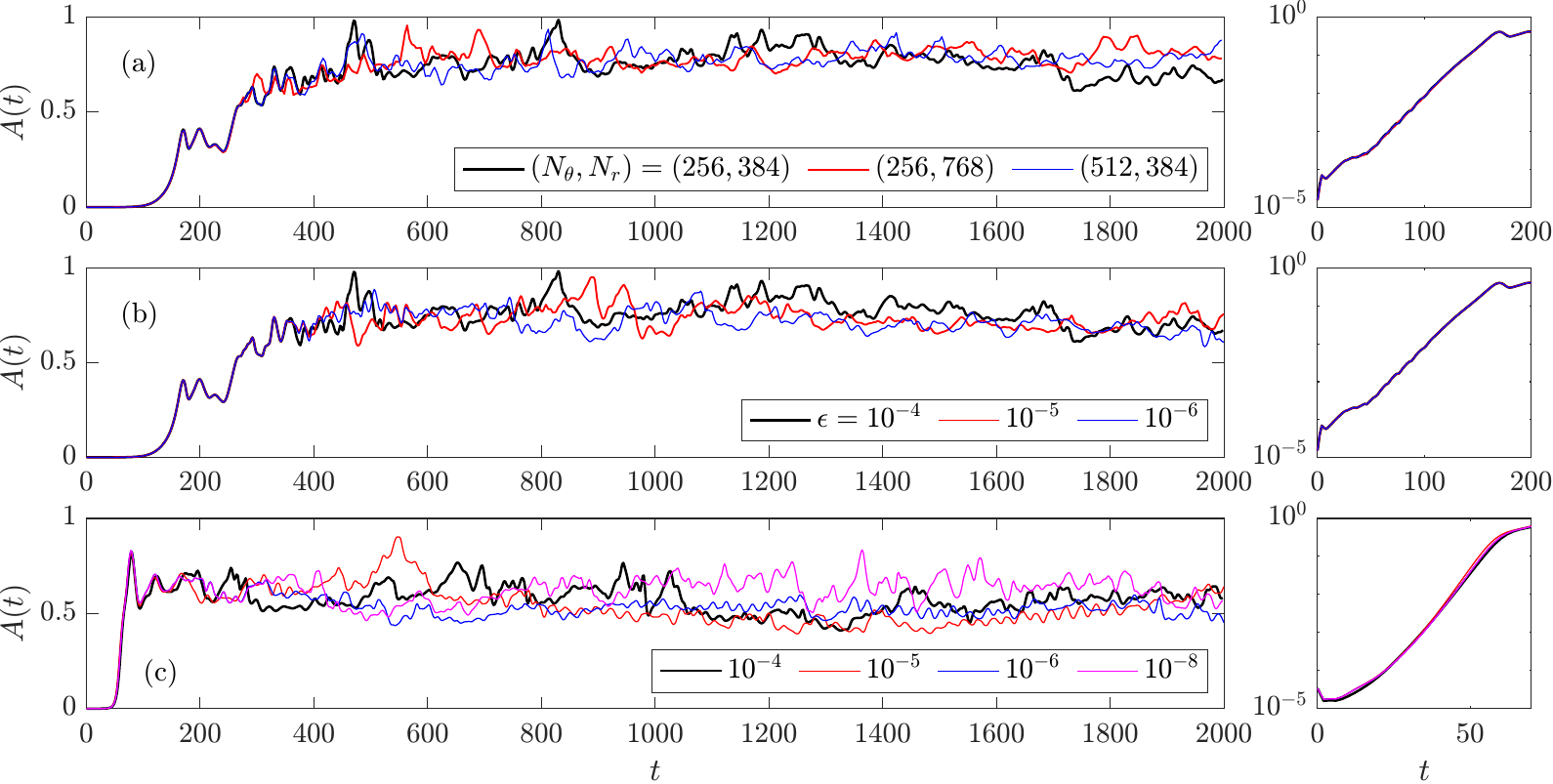}
	\caption{
          (a) Time series of $A(t)$ from simulations with the same parameter values and initial conditions as in figure \ref{fig:tpfig}, with $\epsilon=10^{-4}$ but with different $(N_\theta,N_r)$. The values of $(N_\theta,N_r)$ are indicated in the legend. (b) The same as (a) but with fixed $(N_\theta,N_r)=(256,384)$ and varying $\epsilon$, as indicated in the legend. (c) Time series of $A(t)$ from simulations with the same parameter values and initial conditions as in figure \ref{fig:ppfig}, with fixed $(N_\theta,N_r)=(256,384)$ and varying $\epsilon$. In each row (a)-(c), the right-hand panel shows the same data, but only for early times and on a logarithmic scale. 
          We used a time step of $dt=10^{-3}$ for all simulations with $(N_\theta,N_r)=(256,384)$ but reduced it by half for the simulations with increased $N_\theta$ or $N_r$. In each panel, the thick black line corresponds to exactly the same simulation as is shown in figure \ref{fig:tpfig} or \ref{fig:ppfig}.
          }
	\label{fig:numchecks}
\end{figure}

We typically solve the model equations
using $N_{\theta}=256$ Fourier modes in the $\theta$ direction and $N_r=384$ Chebyshev modes in the $r$ direction, and employ standard 3/2 dealiasing. We use a fixed time step of $dt=10^{-3}$. Figure \ref{fig:numchecks} shows the time series of $A$ from numerical tests where we repeated the simulations from figures \ref{fig:tpfig} and \ref{fig:ppfig} after changing either $(N_\theta,N_r)$ or $\epsilon$. For the higher values of $N_\theta$ or $N_r$ we also halved the time step. In all cases, the 
growth at early times is almost unchanged, and the level at which $A$ saturates is largely unaffected. At later times, there is an inevitable divergence of the $A$ values in the different simulations, since the chaotic dynamics are sensitive to small numerical differences. However, despite each simulation following various trajectories, in each case $A$ continues to oscillate about a similar level. Specifically, for the time series of $A$ in the
simulations in figure \ref{fig:numchecks}, the means $A_{\rm sat}$
and standard deviations $\varsigma(A)$
across the time period $t_{\mathrm{peak}}\leq t< 2000$ are
\be
A_{\rm sat}\pm\varsigma(A) = \{0.732\pm0.117,0.712\pm0.106,0.697\pm0.100\}
\ee
for (a),
\be
A_{\rm sat}\pm\varsigma(A) = \{0.732\pm0.117,0.754\pm0.115,0.742\pm0.111\}
\ee
for (b), and 
\begin{multline}
A_{\rm sat}\pm\varsigma(A) = \{0.546\pm0.097,
0.540\pm0.055,
0.562\pm0.070,0.627\pm0.066\},
\label{Ameansdc}\end{multline}
for (c). The mean value in all these cases is within one standard
deviation of all the others, and there are no clear trends. Hence,
there appears to be no significant impact on the mean level of $A$ 
upon changing the spatial resolution, time step or $\epsilon$. 

The simulations in figure \ref{fig:numchecks}(c) diverge by $t\approx200$, whereas those in figure \ref{fig:numchecks}(a,b) remain close for slightly longer. This difference appears to arise from
the initial conditions, with the stress invariant initially close to
$\Bi$ across a significant portion of the annulus for the simulations in figure \ref{fig:numchecks}(c).
We have also performed tests simulating the full annulus rather than
a periodic quarter domain. 
For these cases, we used $(N_\theta,N_r)=(512,384)$. We again found little
qualitative impact on the dynamics, except that the quarter-domain periodicity was lost at late times as structures with other wavelengths emerged through the growth of round-off errors.

\section{Impact of stress diffusion}\label{app:diff}



\begin{figure}
  \centering
	\includegraphics[width=\linewidth]{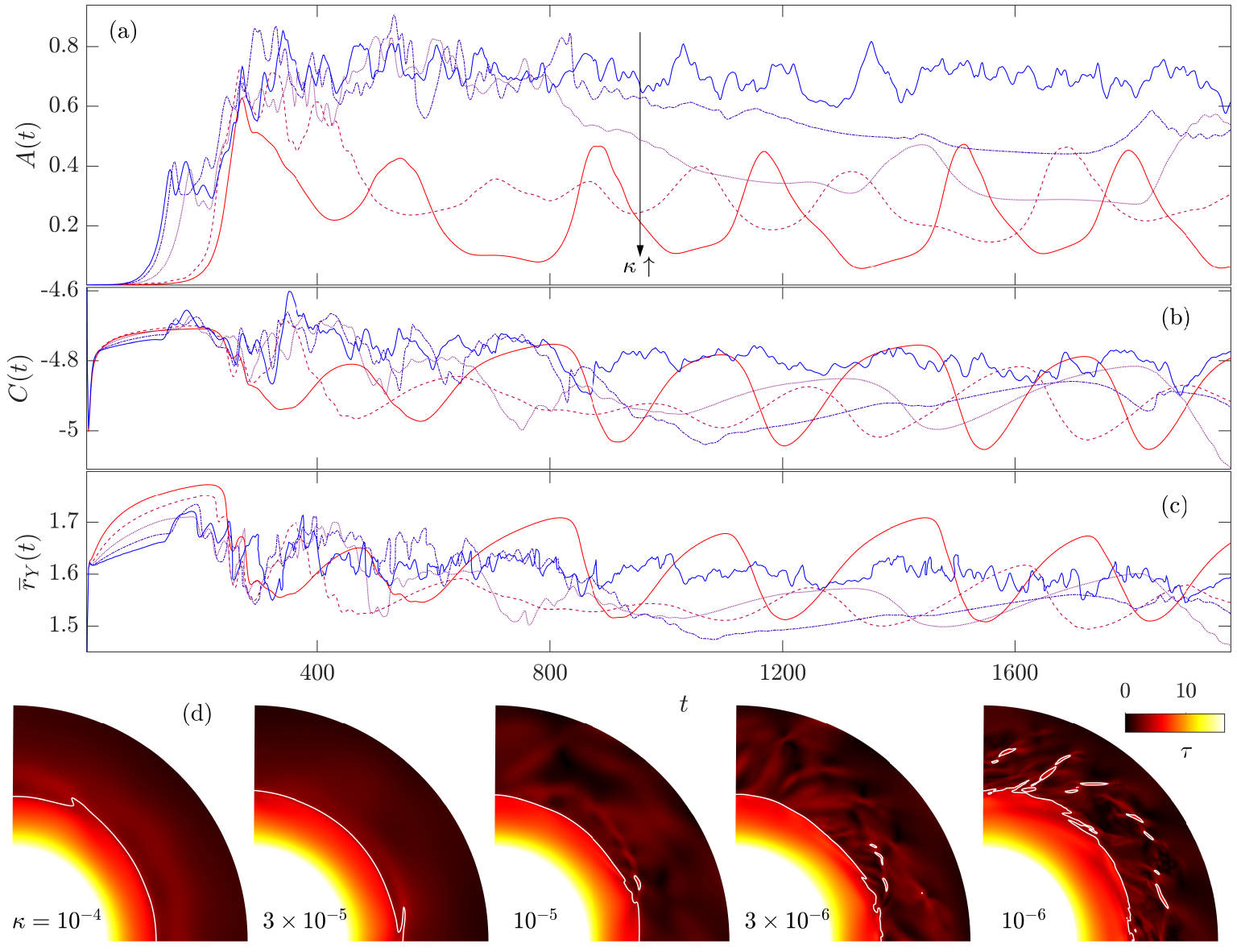}
	\caption{
          Time series of (a) $A(t)$, (b) $C(t)$
          and (c) $\overline{r}_Y(t)$
          for numerical simulations with
          $(\Bi,\Wi,\beta)=(4,1,0.5)$ and
          $\kappa=\{10^{-4},3\times 10^{-5},10^{-5},3\times 10^{-6},10^{-6}\}$
          (colour-coded from red to blue and with the linestyles
          indicated in (a)).
          At the time indicated by the vertical arrow in (a), snapshots of
          $\tau$ are displayed in (d).
          }
	\label{fig:diffig2}
\end{figure}

As reported previously for Poiseuille flow\cite{evturbo},
the nonlinear saturated states are sensitive to diffusion:
so far we have adopted the relatively small value of
$\kappa=10^{-6}$. Raising this diffusivity by up to two
orders of magnitude leads to the results shown in
figure \ref{fig:diffig2}, which displays a suite
of simulations in which, other than $\kappa$, the same parameter
settings are chosen as in the example of figure \ref{fig:tpfig}.
The increase in $\kappa$ slightly decreases
the linear growth rate, such that saturation is attained
at somewhat later times. Note that, based on scaling arguments
and our previous linear stability results for
Poiseuille flow\cite{evturbo}, we expect that modes
become suppressed by diffusion for $m=O(\kappa^{-1/2})$.
Although cut-off is much higher than the $m=4$ wavenumber with
which the simulations are excited in 
figure \ref{fig:diffig2}, the growth rates
are still modified. This modification arises
because a small, but finite polymer
diffusion is capable of lowering stresses
over the pseudo-plug below $\Bi$, which adjusts the switch
in \eqref{eq:LSA_fhat} ({\it cf.} Ref.~\onlinecite{evturbo}).

Despite the difference in linear growth rate, the saturation levels
first reached
are similar across the different diffusivities;
see figure \ref{fig:diffig2}(a) for $t=O(100)$.
Further beyond saturation, however, polymer diffusion leads to 
the amplitude $A(t)$ waning over longer times. Simultaneously, the
non-axisymmetrical spatial
pattern becomes smoother and its temporal variability becomes
more regular (see the snapshots also displayed in
figure \ref{fig:diffig2}). Even relatively low polymer diffusion
therefore exerts a pervasive effect on the nonlinear states. 
For $\kappa=10^{-6}$, $A(t)$ maintains a level close to that at which it first saturates for the duration of the simulation. This suggests that diffusion is not having a significant effect for that case, so robust measurements of the nonlinear dynamics can be made.

Although the more diffusive examples in figure \ref{fig:diffig2}
are less complicated, the time dependence of the solution
nevertheless reveals interesting dynamics: diffusion
smooths the solutions but does not
eliminate instability.
As a result, unstable modes appear sporadically
in a sequence of roughly regular bursting events.
We have not explored this diffusive dynamics in any detail, however,
and examine computations with $\kappa=10^{-6}$ in the main text.

\section{Short-wavelength dispersion relation}\label{sec:appB}

From \eqref{eq:ry_explicit} it follows that
\be
\trt0(r_p) = \frac{C_\infty}{r_p^2} = -
\sqrt2 \Bi \left[1 + \sqrt{1 + \frac{4\Wi^2\Bi^2}{(1-\beta)^2}}\right]^{-1/2}
\equiv \Bi \frt(\Wi\Bi,\beta)
.
\ee
Moreover,
\be
\ttt0(r_p) = \Bi \ftt(\Wi\Bi,\beta),
\quad
\ftt = 2\sqrt{1-\frt^2}.
\ee
The short-wave dispersion relation for the local radial wavenumber $k$
at the border of the
pseudo-plug $r=r_p$ can then be reduced to
\be
D(k) = c_0 + c_1 K + c_2 K^2 + c_3 K^3 + c_4 K^4 = 0,
\ee
where
\be
c_0 = \beta \Wi\Bi \Sigma^2 + \Sigma
+ \ftt ( 1+\Wi\Bi\Sigma) 
,\qquad
c_1 = 2\frt ( \Wi\Bi\Sigma + 3),
\ee
\be
c_2 =  2\beta\Wi\Bi\Sigma^2 +
\Wi\Bi\ftt\Sigma  + 2\Sigma - \ftt
+ 4(\Wi\Bi)^{-1}(1-\beta),
\ee
\be
c_3 = 2\frt (\Wi\Bi\Sigma - 1)
  ,\qquad
  c_4 = \Sigma(\beta\Wi\Bi\Sigma + 1)
  ,
\ee
\be
\sigma = \Bi \Sigma
\qquad {\rm and} \qquad
K = r_p k.
\ee
The dispersion relation has the special feature that
pairs of the local wavenumbers merge into double roots
for a particular value of $\Sigma$. {\it i.e.}
\be
D(K) = c_4 (K^2 + a K + b)^2,
\ee
with
\be
a = \frac{c_3}{2c_4}
\qquad {\rm and} \qquad
b = \frac{c_1}{c_3}.
\ee
The condition that demands that there is a pair of double roots
then boils down to
\be
64c_4^3c_0 - (c_3^2 - 4c_2c_4)^2 =
8c_4^2c_1 + c_3(c_3^2 - 4c_2c_4) = 0
,
\label{C.8}
\ee
or
\be
c_4 c_1^2 - c_0 c_3^2  =
4 \frt^2 (1+\Wi\Bi\Sigma) \left[\Wi\Bi (8\beta - \ftt \Wi \Bi) \Sigma^2
  + 2 (4+\ftt\Wi\Bi )\Sigma - \ftt \right] = 0
\label{C.11}
.
\ee
In other words,
\be
\sigma = \Bi \Sigma(\Wi\Bi,\beta),
\ee
where
\be
\Sigma = \frac{4+\ftt\Wi\Bi - 2\sqrt{4 + 2\ftt \Wi\Bi (1+\beta)}}
{\Wi\Bi(\ftt\Wi\Bi-8\beta)},
\ee
which leads to two real pairs of local wavenumbers (the other 
solution to \eqref{C.11} for $\Sigma$ that is positive does not).

\section{Smaller $\beta$ in nonlinear simulations}\label{app:beta}

\begin{figure}
  \centering
	\includegraphics[width=\linewidth]{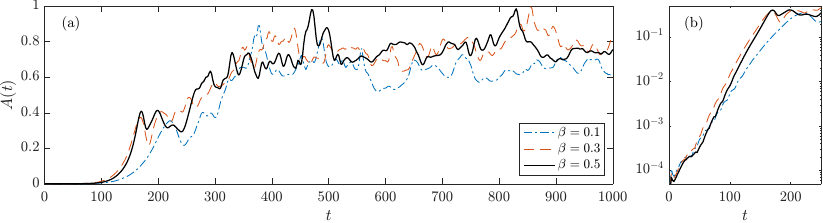}
	\caption{
          (a) Time series of $A(t)$ for simulations with $\{\Wi,\Bi,R\}=\{1,4,\frac52\}$, initial stress condition \eqref{stressIC}, and $\beta\in\{0.1,0.3,0.5\}$. (b) The same as in (a) but plotted on a logarithmic scale and with only early times shown. 
          }
	\label{fig:bevar}
\end{figure}

To test the impact of decreasing $\beta$ on the nonlinear dynamics, we ran numerical simulations similar to those in figure \ref{fig:tpfig}, but with $\beta=0.1$ and $\beta=0.3$. The results for $A(t)$ are shown in figure \ref{fig:bevar}, along with the same quantity for the simulation from figure \ref{fig:tpfig} with $\beta=0.5$. There is a modest impact of changing $\beta$ on the early-time growth (panel (b)). Some impact of $\beta$ on the growth rate is expected from linear theory (\S\ref{sec:LSshort}), although due to the polymer-stress-free initial conditions used in these simulations, the normal-mode theory is not directly applicable. In any case, $A(t)$ saturates at a similar value in each of the simulations and is sustained around that level for a significant time (panel (a)). This indicates that varying the viscosity ratio within the range $0.1\leq\beta\leq 0.5$ does not have a significant impact on the level of complexity of the nonlinear dynamics generated via linear instability. 

\end{appendix}

\bibliographystyle{elsarticle-num}
\bibliography{../../../vpthread/jfm.bib}

\end{document}